\documentclass[
  prd,
  preprint,
  tightenlines,
  superscriptaddress,
  preprintnumbers,
  nofootinbib
]{revtex4}
\usepackage{graphicx}
\usepackage{epsfig}
\usepackage{comment}
\usepackage{bm}
\usepackage{latexsym,amssymb,amsmath,wasysym,float}

\usepackage{color}

\usepackage{scrextend}
\usepackage{mathrsfs}
\usepackage{enumitem}

\usepackage[usenames,dvipsnames]{xcolor}
\definecolor{orange}{cmyk}{0,0.5,1,0}
\definecolor{rossoCP3}{cmyk}{0,.88,.77,.40}
\definecolor{graa}{rgb}{0.8,0.8,0.8}
\definecolor{blaa}{rgb}{0.2,0.2,0.6}

\definecolor{darkgreen}{rgb}{0.,0.6,0.}

\makeatletter
\g@addto@macro\frontmatter@RRAP@format{\vspace{6pt}}
\g@addto@macro\ps@titlepage{
  \let\@oddfoot\@empty
  \let\@evenfoot\@empty
}
\AtBeginDocument{
	\renewcommand{\frontmatter@title@below}{\vspace{10pt}}
	\renewcommand{\frontmatter@preabstractspace}{35pt}
  	\renewcommand{\frontmatter@affiliationfont}{
		\def\baselinestretch{1}
		\small\itshape
		\setlength{\parskip}{3pt}
	}
}
\makeatother

\renewcommand{\baselinestretch}{1.2}

\usepackage[colorlinks=true,linktoc=all,linkcolor=blaa,citecolor=blaa,urlcolor=blaa]{hyperref}

\begin{document}
\enlargethispage{3\baselineskip}

\preprint{DESY-26-147}
\preprint{MPP-2026-204}


\title{\color{rossoCP3}
Quantum Cosmology at the Cusp
}

\author{\bf Luis A. Anchordoqui}

\affiliation{Department of Physics and Astronomy,  Lehman College, City University of
  New York, NY 10468, USA 
}

\affiliation{Department of Physics,
 Graduate Center,  City University of
  New York,  NY 10016, USA 
}

\affiliation{Department of Astrophysics,
 American Museum of Natural History, NY
 10024, USA
}

\author{\bf Alek Bedroya}

\affiliation{Leinweber Forum for Theoretical Physics, Princeton University, Princeton, NJ 08544, USA }

\author{\bf Muldrow\nolinebreak~Etheredge}

\affiliation{Max--Planck--Institut f\"ur Physik,
 Werner--Heisenberg--Institut, 85748 Garching, Germany }

\author{\bf Dieter\nolinebreak~L\"ust}

\affiliation{Max--Planck--Institut f\"ur Physik,
 Werner--Heisenberg--Institut, 85748 Garching, Germany }

\affiliation{Arnold Sommerfeld Center for Theoretical Physics,
Ludwig-Maximilians-Universit\"at M\"unchen,
80333 M\"unchen, Germany
}

\author{\bf Severin L\"ust}
\affiliation{Sorbonne Universit\'e, CNRS, Laboratoire de Physique Th\'eorique et Hautes \'Energies, LPTHE, F-75005 Paris, France }

\author{\bf Alexander Westphal}
\affiliation{ Deutsches Elektronen-Synchrotron DESY, Notkestr.~85, 22607 Hamburg, Germany }

\date{\today}

\begin{abstract}
In top-down quantum gravity constructions, saxionic directions typically lead to cusps in moduli space, where the scalar potential and the axion decay constant asymptotically behave as $V\sim e^{-\alpha\phi}$ and $f\sim e^{-\gamma\phi}$ in reduced Planck units. Such cusps admit axion-supported scaling cosmologies that expand faster than the corresponding solutions with the axion held fixed, for the same scalar potential. In spatially compact universes, axionic momentum is quantized, and any semiclassical state in the cusp with a localized axion carries nonzero axionic kinetic energy. We show using the Wheeler-DeWitt equation that this contribution drives the late-time evolution towards the axion-supported scaling solution whenever $(d-2)\alpha(\alpha+2\gamma)>4(d-1)$ in $d$ spacetime dimensions. We derive an upper bound on
the transition time. Long-lived accelerated expansion requires $2\gamma>(d-2)\alpha$. The trans-Planckian censorship conjecture requires the opposite inequality, $2\gamma\leq(d-2)\alpha$, which also follows independently from requiring the brane magnetically charged under the axion to remain heavier than the Hubble scale. 
\end{abstract}

\maketitle


\tableofcontents

\section{Introduction}
\label{sec:intro}

Scalar fields and their moduli spaces play a central role in string compactifications and in the low-energy description of quantum gravity. A particularly important lesson of the Swampland program is that the geometry of these scalar field spaces is highly constrained, especially near infinite-distance limits \cite{Vafa:2005ui,Ooguri:2006in}. In such limits, infinite towers of states become light, approximate continuous symmetries can emerge, and many physical quantities exhibit universal exponential behavior \cite{Lee:2019wij,Etheredge:2022opl,Castellano:2023stg, Castellano:2023jjt,Etheredge:2024tok, Etheredge:2025ahf,Aoufia:2026ztl,Etheredge:2026cha}. Recent work involving the Emergent String Conjecture \cite{Lee:2019wij} has shown that these asymptotic structures can be organized in terms of discrete lattices of exponential decay vectors associated with towers, branes, the species scale, axion decay constants, and scalar potentials \cite{Etheredge:2025ahf,Aoufia:2026ztl,Etheredge:2026cha}.

In this work we focus on the asymptotic geometry of axionic directions in quantum gravity. Consider a periodic scalar
\begin{equation}
\theta\simeq\theta+2\pi\, .
\end{equation}
The corresponding orbit defines a non-contractible cycle unless it shrinks at a fixed point. The conjectures of Ooguri and Vafa imply that such cycles cannot retain a strictly positive minimum circumference throughout the complete moduli space \cite{Ooguri:2006in}. In the known infinite-distance limits of string theory, this shrinking is realized through cusp-like geometries. Non-perturbative effects that break the continuous axionic shift symmetry become exponentially suppressed, and the continuous shift symmetry is restored asymptotically \cite{Lanza:2021eft}. Recent studies of asymptotic moduli-space geometry make this structure particularly explicit and show that axionic fibers generically shrink exponentially over the non-compact saxionic directions \cite{Aoufia:2026ztl}.

We first review the axion-supported scaling cosmologies that arise in such cusps, which give rise to faster expansion for the same scalar potential due to the non-trivial coupling between the axion and the saxion. The axionic shift symmetry implies a conserved momentum $J$, and for an axion decay constant $f^2=f_0^2e^{-2\gamma\phi}$ the corresponding classical kinetic energy density takes the form
\begin{equation}
\rho_\theta={J^2\over2f_0^2}a^{-6}e^{2\gamma\phi}\, .
\end{equation}
The explicit factor $a^{-6}$ tends to dilute the axion kinetic energy, while the shrinking decay constant exponentially enhances it as the saxion $\phi$ moves towards the cusp. In the presence of an exponential potential
\begin{equation}
V(\phi)=V_0e^{-\alpha\phi}\, ,
\end{equation}
these two effects can balance each other and give scaling cosmologies. These solutions were found in Refs.~\cite{Christodoulidis:2019jsx,Revello:2023hro}, and we refer to them as axion-supported scaling solutions. For a large turn rate of the field-space trajectory, they correspond to the rapid-turn solutions of Refs.~\cite{Brown:2017hyper,Bjorkmo:2019rapid,Mizuno:2017idt,Bjorkmo:2019aev}, and they have been studied in the context of stringy quintessence in Refs.~\cite{Seo:2024qzf,Grimm:2025cpq}. In general $d$ spacetime dimensions, the exact axion-supported scaling solution has (see Sec.~\ref{sec:scaling-solutions} and Ref.~\cite{Revello:2023hro})
\begin{equation}
a(t)\propto t^{(1+\beta)/(d-1)}\, ,\qquad \beta={2\gamma\over\alpha}\, ,
\end{equation}
and equation of state
\begin{equation}
w={1-\beta\over1+\beta}\, .
\end{equation}
Whenever it exists, this solution expands faster than the tracker solution $a(t)\propto t^{4/((d-2)\alpha^2)}$, which is driven by the same potential in the absence of the axion, and than kination, $a(t)\propto t^{1/(d-1)}$, which is the late-time solution for steeper potentials.

Our main new observation is that, quantum mechanically, these solutions are universal when spatial slices are compact, and we derive an upper bound on how long it takes to reach them.
In previous work \cite{Anchordoqui:2025izb,Anchordoqui:2026nit}, some of the authors studied another aspect of the asymptotic structure of moduli spaces by treating the moduli themselves quantum mechanically. The natural Hamiltonian of this quantum mechanics contains the Laplacian on moduli space,
\begin{equation}
{\cal H}_{\rm matt}=-{1\over2}\Delta_{\cal M}+V\, .
\end{equation}
In the present work we couple this system to gravity. Restricting to the homogeneous and isotropic sector, the gravitational degree of freedom is the scale factor $a$, or equivalently the overall spatial volume, and the combined system of gravity and moduli $\phi^i$ is described by the Wheeler-DeWitt equation \cite{Wheeler:1968iap,DeWitt:1967yk,Gielen:2023hsg} (see Sec.~\ref{sec:WDWeq})
\begin{equation}
\left({\cal H}_{\rm grav}+{\cal H}_{\rm matt}\right)\Psi(a,\phi^i)=0\, .
\end{equation}
For four-dimensional Friedmann-Lema\^{\i}tre-Robertson-Walker (FLRW) cosmology with spatial slices of unit comoving volume, the matter Hamiltonian takes the form (see Sec.~\ref{sec:vanishingV})
\begin{equation}
{\cal H}_{\rm matt}(a,\phi^i)=-{1\over 2a^3}\Delta_{\cal M}+a^3V(\phi^i)\, .
\end{equation}
The factor $a^{-3}$ has an important physical meaning, and the cusp geometry makes the corresponding contribution, in some cases, unavoidable for semiclassical cosmological states.

A finite-volume moduli space may possess a constant normalizable zero mode. However, this state is completely delocalized and cannot describe a semiclassical saddle in which the scalar fields take approximately definite values. Since the cusp has a positive asymptotic spectral threshold, a wave packet localized sufficiently far inside the cusp necessarily contains Laplacian modes with non-zero eigenvalue. If the axion itself is localized, the state must additionally contain non-zero Fourier modes $n\neq0$, whose contribution becomes increasingly important as the axion circle shrinks. This observation has a direct cosmological implication. The Laplacian produces a comoving energy proportional to $a^{-3}$ and hence an effective energy density (see Eq.~\eqref{eq:universal-a6})
\begin{equation}
\rho_{\rm q}={\lambda_\psi\over 2a^6}\, ,
\end{equation}
where $\lambda_\psi$ is the expectation value of $-\Delta_{\cal M}$ in the state $\psi$, which is constant in the absence of a potential and otherwise evolves with the state. Therefore, even if this contribution is negligible at large spatial volume, it grows rapidly towards the past. The quantum-mechanical energy required to localize a classical configuration on a cuspy moduli space thus generically becomes cosmologically relevant at sufficiently small $a$. In $d$ spacetime dimensions the corresponding contribution scales as $a^{-2(d-1)}$.

Moreover, on compact spatial slices, the axionic component of this energy is quantized. Classically, the conserved angular momentum can be taken arbitrarily close to zero, whereas quantum mechanically a semiclassical state in which the axion is localized within an interval $\Delta\theta$ necessarily has a root-mean-square axionic momentum $n_{\rm rms}\geq1/\Delta\theta$ (see Eq.~\eqref{eq:nrms}). Consequently, the onset of the axion-supported scaling regime cannot be postponed arbitrarily by continuously tuning the angular momentum to zero. Starting from the scalar-dominated exponential scaling solution, the relative importance of a non-zero axion momentum grows whenever (see Eq.~\eqref{eq:rapidturn-existence-d})
\begin{equation}
\alpha(\alpha+2\gamma)> \frac{4(d-1)}{d-2} \,,
\end{equation}
where $\alpha$ and $\gamma$ denote the exponential slopes of the scalar potential and the axion decay constant, as introduced above.
Under this condition the tracker solution driven by the potential is unstable to any non-zero axionic momentum \cite{Christodoulidis:2019jsx,Revello:2023hro}, and the system is eventually driven towards the axion-supported scaling attractor. In closed universes, the axion-supported scaling solution therefore dominates the future evolution of any semiclassical state in which the axion is localized.

We also find an upper bound on the time-scale of this transition. We show that this happens after a number of $e$-folds that depends only logarithmically on the initial fractions of the energy density in the potential and in the axion kinetic energy, and through the latter on the initial spatial volume, whether the energy density is initially dominated by the saxion kinetic energy, the axion kinetic energy, or the potential (see Eqs.~\eqref{eq:NRT-bound} and \eqref{eq:NRT-quantum}). The spectral structure of the cusp and the quantization of the axion momentum therefore determine not only the existence of quantum corrections, but also when they must become dynamically relevant.

Finally, one may ask whether the cusp region can give rise to accelerated expansion. Axion-supported scaling solutions are particularly interesting from the viewpoint of the Swampland because negatively curved field spaces can support accelerated expansion even when the scalar potential alone is too steep to do so \cite{Achucarro:2018vey,Akrami:2020zfz}. For the axion-supported scaling solution, eternal acceleration requires  $2\gamma>(d-2)\alpha$. 

We could not realize axion-supported eternal accelerated expansion in top-down constructions. We find that in decompactification limits, the allowed axion decay rates and potential slopes exclude eternal acceleration. The same is true in perturbative emergent-string limits, where the allowed values of $\gamma$ and $\alpha$ are constrained by the string perturbation expansion and the taxonomy of asymptotic branes and potentials \cite{Aoufia:2026ztl,Etheredge:2025ahf,Etheredge:2026cha}.

Eternally accelerating power-law cosmologies would violate the trans-Planckian censorship conjecture (TCC) \cite{Bedroya:2019snp}. Although TCC remains a conjecture, it can be derived from holography under certain assumptions in special cases, such as the future-infinity limit of interest here \cite{Bedroya:2022tbh,Bedroya:2024zta}. The TCC therefore implies a new asymptotic inequality for axions 
\begin{equation}
2\gamma\leq(d-2)\alpha\, .
\end{equation}
We show that this inequality follows from the taxonomy rules which provides a non-trivial test of the TCC.

Note that axion-supported scaling cosmologies can violate the TCC even when the scalar potential satisfies the strong de Sitter conjecture $|\nabla V|/V\geq2/\sqrt{d-2}$ \cite{Rudelius:2021oaz,Rudelius:2021azq}. This shows that, at future infinity, TCC is a stronger condition, since it constrains the solutions rather than the scalar potential alone.

After finding the bound $2\gamma\leq(d-2)\alpha$ from the TCC, and while writing this paper, we became aware of Ref.~\cite{Collazuol:2026xuz}, which in parallel finds an independent argument for it in four dimensions, based on requiring the tension scale of EFT strings to remain above the Hubble scale. In Sec.~\ref{sec:EFT-strings} we build on their work and generalize this argument to arbitrary spacetime dimensions. We show that requiring the brane magnetically charged under the axion to remain heavier than the Hubble scale reproduces exactly the inequality $2\gamma\leq(d-2)\alpha$ in $d$ spacetime dimensions.

For other approaches that aim to obtain accelerated expansion, in particular de Sitter space as an excited state, which differ from the approach taken here, (see Refs.~\cite{Goheer:2002vf,Dvali:2014gua,Maltz:2016iaw,Dvali:2017eba,Dvali:2018fqu,Dvali:2018jhn}).


The remainder of the paper is organized as follows. In Sec.~\ref{sec:scaling} we study the classical scaling cosmologies of an axion-saxion system at a cusp in the presence of an exponential potential. We derive the axion-supported scaling solution in Sec.~\ref{sec:scaling-solutions}, and in Sec.~\ref{sec:transition} we describe the kination, tracker, and axion-supported solutions and bound the number of $e$-folds it takes to reach the axion-supported regime for different initial conditions. In Sec.~\ref{sec:quantum} we turn to quantum effects. In Sec.~\ref{sec:WDW} we review the reduction of gravity coupled to moduli to the homogeneous Wheeler-DeWitt system and explain the role of the spatial volume modulus, and in Sec.~\ref{sec:signature} we comment on the Lorentzian signature of the combined gravity-plus-moduli kinetic term and on its relation to the species frame. We then study the moduli-space Laplacian and its scale-factor dependence, first without (Sec.~\ref{sec:vanishingV}) and then with (Sec.~\ref{sec:nonvanishingV}) a scalar potential. In Sec.~\ref{sec:localized} we analyze localized wave functions near asymptotic cusps, show that on compact spatial slices they carry a non-zero root-mean-square axionic momentum, and derive an upper bound on the number of $e$-folds before the axion-supported regime sets in. In Sec.~\ref{sec:TCC} we use the TCC to derive a new inequality for axions, Eq.~\eqref{eq:TCC-rapidturn}, contrast it with the strong de Sitter conjecture, and test it against the taxonomy rules for asymptotic branes and potentials and the asymptotic data of standard string-theory limits, and in Sec.~\ref{sec:EFT-strings} we relate it to light EFT strings and to recent results on asymptotic accelerated expansion in four-dimensional ${\cal N}=1$ theories. We summarize our results in Sec.~\ref{sec:summary}. Appendix~\ref{sec:appA} compares the field-momentum and field-velocity commutators in an FLRW background, and Appendix~\ref{sec:appB} discusses the resulting scale-factor dependence of the Emergent String Conjecture-related taxonomy rules \cite{Castellano:2023stg, Castellano:2023jjt,Etheredge:2024tok, Etheredge:2025ahf,Aoufia:2026ztl,Etheredge:2026cha}.

\section{Scaling cosmologies at cusps}
\label{sec:scaling}

Consider a periodic scalar $\theta\simeq\theta+2\pi$. The corresponding orbit defines a non-contractible cycle unless it shrinks at a fixed point. The conjectures of Ooguri and Vafa imply that such cycles cannot retain a strictly positive minimum circumference throughout the complete moduli space \cite{Ooguri:2006in}. In the known infinite-distance limits of string theory, this shrinking is realized through cusp-like geometries. Non-perturbative effects that break the continuous axionic shift symmetry become exponentially suppressed, and the continuous shift symmetry is restored asymptotically \cite{Lanza:2021eft}. Recent studies of asymptotic moduli-space geometry make this structure particularly explicit and show that axionic fibers generically shrink exponentially over the non-compact saxionic directions \cite{Aoufia:2026ztl}. In this section we study the classical cosmological evolution of such an axion-saxion system in the presence of an exponential potential, and show that a non-zero axionic momentum eventually drives the system towards an axion-supported scaling cosmology.

\subsection{Scaling cosmologies}
\label{sec:scaling-solutions}

It is useful to work in general $d$ spacetime dimensions, in reduced Planck units, and to consider the two-field system of a saxion $\phi$ and an axion $\theta\simeq\theta+2\pi$ with action
\begin{equation}
S=\int d^dx\sqrt{-g}\,\Bigl[\frac{1}{2}R-\frac{1}{2}(\partial\phi)^2-\frac{1}{2}f(\phi)^2(\partial\theta)^2-V(\phi)\Bigr]\, ,
\label{eq:axion-saxion-action}
\end{equation}
where the axion decay constant and the scalar potential behave asymptotically as
\begin{equation}
f(\phi)=f_0e^{-\gamma\phi}\, ,\qquad V(\phi)=V_0e^{-\alpha\phi}\, ,
\label{eq:general-rapidturn}
\end{equation}
with $\gamma,\alpha,V_0>0$. The corresponding moduli-space metric, $ds_{\cal M}^2=d\phi^2+f_0^2e^{-2\gamma\phi}d\theta^2$, describes a cusp in which the axion circle shrinks exponentially as one approaches the infinite-distance limit $\phi\rightarrow+\infty$. We assume that the potential depends only on $\phi$ and does not break the shift symmetry of $\theta$. In string theory, we can, for example, identify $\phi$ with the canonically normalized dilaton, in which case $V$ corresponds to a perturbative potential scaling as a power of the string coupling. Another option is that $\phi$ corresponds to a geometric modulus, i.e. the logarithm of a radius $R$, and then $V$ scales as a negative power of $R$.

We consider a spatially flat FLRW metric $ds^2=-dt^2+a(t)^2d\vec x^{\,2}$ and homogeneous fields $\phi(t)$ and $\theta(t)$. The Friedmann equations read
\begin{equation}
\frac{(d-1)(d-2)}{2}H^2=\frac{1}{2}\dot\phi^2+\frac{1}{2}f(\phi)^2\dot\theta^2+V(\phi)\, ,\qquad (d-2)\dot H=-\dot\phi^2-f(\phi)^2\dot\theta^2\, ,
\label{eq:friedmann}
\end{equation}
where $H=\dot a/a$, and the equations of motion of the saxion and of the axion are
\begin{equation}
\ddot\phi+(d-1)H\dot\phi+\gamma f(\phi)^2\dot\theta^2-\alpha V(\phi)=0\, ,\qquad \frac{d}{dt}\Bigl(a^{d-1}f(\phi)^2\dot\theta\Bigr)=0\, ,
\label{eq:eom}
\end{equation}
where we used $f'=-\gamma f$ and $V'=-\alpha V$. The potential pushes the saxion towards the cusp, whereas the axion kinetic term pushes it away from the cusp.

The shift symmetry of $\theta$ implies the conservation of
\begin{equation}
J=a^{d-1}f_0^2e^{-2\gamma\phi}\dot\theta\, ,
\label{eq:Jconserved}
\end{equation}
which is the axionic momentum per unit comoving volume. For compact spatial slices with comoving volume $V_c$, the total axionic momentum, i.e. the momentum canonically conjugate to the zero mode of $\theta$, is $V_cJ$. This will play an important role in Sec.~\ref{sec:localized}. Conservation of $J$ allows us to eliminate the axion velocity, $\dot\theta=J/(a^{d-1}f^2)$, and the axion kinetic energy density becomes
\begin{equation}
\rho_\theta\equiv\frac{1}{2}f(\phi)^2\dot\theta^2=\frac{J^2}{2f_0^2}a^{-2(d-1)}e^{2\gamma\phi}\, .
\label{eq:rho-theta}
\end{equation}
The explicit factor $a^{-2(d-1)}$ tends to dilute the axion kinetic energy, while the shrinking decay constant exponentially enhances it as the saxion $\phi$ moves towards the cusp. The system therefore reduces to a single field $\phi$ coupled to gravity,
\begin{equation}
\frac{(d-1)(d-2)}{2}H^2=\frac{1}{2}\dot\phi^2+V_{\rm eff}(\phi,a)\, ,\qquad \ddot\phi+(d-1)H\dot\phi+\partial_\phi V_{\rm eff}(\phi,a)=0\, ,
\label{eq:single-field}
\end{equation}
with the effective potential
\begin{equation}
V_{\rm eff}(\phi,a)=V_0e^{-\alpha\phi}+\frac{J^2}{2f_0^2}a^{-2(d-1)}e^{2\gamma\phi}\, ,
\label{eq:Veff}
\end{equation}
where $\partial_\phi$ is taken at fixed $a$. Note that $\dot\theta$ has to be eliminated in the equations of motion, or equivalently in the Routhian at fixed $J$, rather than by direct substitution into the action. The axionic contribution then enters $V_{\rm eff}$ with a positive sign and acts as a centrifugal barrier that prevents the saxion from running into the cusp. Since the exponential potential pushes the saxion towards the cusp, the two terms in $V_{\rm eff}$ can balance each other and give scaling cosmologies. These scaling solutions were found in Ref.~\cite{Christodoulidis:2019jsx} in four dimensions and in Ref.~\cite{Revello:2023hro} for saxion-axion systems with the kinetic structure of four-dimensional ${\cal N}=1$ supergravity in general $d$, and we review them below. We refer to them as axion-supported scaling solutions. Since the non-zero axion velocity makes the field-space trajectory turn continuously, they are related to the rapid-turn solutions of Refs.~\cite{Brown:2017hyper,Bjorkmo:2019rapid,Mizuno:2017idt,Bjorkmo:2019aev}, although the turn rate is not necessarily large, see below.

We look for a scaling solution of the form
\begin{equation}
a(t)\propto t^p\, ,\qquad \phi(t)=\phi_0+\frac{2}{\alpha}\log t\, ,
\label{eq:scaling-ansatz}
\end{equation}
for which the potential scales as $t^{-2}$. The axion kinetic energy \eqref{eq:rho-theta} also scales as $t^{-2}$ provided
\begin{equation}
p=\frac{1+\beta}{d-1}\, ,\qquad \beta=\frac{2\gamma}{\alpha}\, .
\label{eq:p-beta-d}
\end{equation}
Writing $\rho_\theta=x_{\rm AS}\,t^{-2}$ and $V=y_{\rm AS}\,t^{-2}$, the saxion equation of motion gives $y_{\rm AS}=\beta x_{\rm AS}+2\beta/\alpha^2$, and the Friedmann equation then fixes
\begin{equation}
x_{\rm AS}=\frac{(d-2)(1+\beta)\alpha^2-4(d-1)}{2(d-1)\alpha^2}\, ,\qquad y_{\rm AS}=\frac{(d-2)\beta(1+\beta)}{2(d-1)}\, .
\label{eq:xyRT}
\end{equation}
The explicit axion-supported scaling solution is therefore
\begin{equation}
a(t)=\biggl[\frac{J^2}{2f_0^2x_{\rm AS}}\Bigl(\frac{V_0}{y_{\rm AS}}\Bigr)^{\beta}\biggr]^{\frac{1}{2(d-1)}}t^{\frac{1+\beta}{d-1}}\, ,\qquad \phi(t)=\frac{1}{\alpha}\log\Bigl(\frac{V_0t^2}{y_{\rm AS}}\Bigr)\, ,
\label{eq:RT-solution}
\end{equation}
where $t$ is measured from the initial singularity. The axion momentum $J$ only sets the overall normalization of the scale factor, as expected from the symmetry $a\rightarrow ca$, $J\rightarrow c^{d-1}J$. On this solution the fractions of the energy density $\rho=(d-1)(d-2)H^2/2$ in the potential, in the axion kinetic energy, and in the saxion kinetic energy are constant,
\begin{equation}
\Omega_V=\frac{\beta}{1+\beta}\, ,\qquad \Omega_\theta=\frac{(d-2)(1+\beta)\alpha^2-4(d-1)}{(d-2)\alpha^2(1+\beta)^2}\, ,\qquad \Omega_{\dot\phi}=\frac{4(d-1)}{(d-2)\alpha^2(1+\beta)^2}\, ,
\label{eq:Omega-RT}
\end{equation}
and the ratio of axion kinetic energy to potential energy is
\begin{equation}
{\cal R}_{\rm AS}\equiv\frac{\rho_\theta}{V}\bigg|_{\rm AS}=\frac{(d-2)(1+\beta)\alpha^2-4(d-1)}{(d-2)\alpha^2\beta(1+\beta)}\, .
\label{eq:RRT}
\end{equation}
Since $\rho_\theta$ must be positive, a non-zero axion-supported branch exists when
\begin{equation}
\alpha^2(d-2)(1+\beta)>4(d-1)\, .
\label{eq:rapidturn-existence-d}
\end{equation}
At the boundary of this region $\rho_\theta\rightarrow0$, and the axion-supported solution merges with the tracker solution discussed in Sec.~\ref{sec:transition}. The axion-supported scaling solution is the fixed point of Ref.~\cite{Revello:2023hro} at which the axion moves, which is stable whenever it exists, and for $d=4$ it reduces to the scaling solution of Ref.~\cite{Christodoulidis:2019jsx}, where it emerges from the tracker solution through a bifurcation. The corresponding equation of state is
\begin{equation}
w=\frac{1-\beta}{1+\beta}=\frac{\alpha-2\gamma}{\alpha+2\gamma}\, ,
\label{eq:w-beta-d}
\end{equation}
and acceleration requires
\begin{equation}
p>1\qquad\Longleftrightarrow\qquad\beta>d-2\, .
\label{eq:acceleration-d}
\end{equation}
In four dimensions, the axion-supported scaling solution has $a\propto t^{(1+\beta)/3}$, exists for $\alpha^2(1+\beta)>6$, and is accelerating for $\beta>2$, i.e. $\gamma>\alpha$. Finally, the covariant turn rate of the field-space trajectory in units of the Hubble rate is $\gamma\sqrt{(d-1)(d-2)\Omega_\theta}$. The trajectory therefore turns rapidly only if $\gamma^2(d-1)(d-2)\Omega_\theta\gg1$, and the turn rate vanishes at the boundary of the existence region \eqref{eq:rapidturn-existence-d}, where $\Omega_\theta\rightarrow0$.

\subsection{Transition time-scale}
\label{sec:transition}

It is useful to estimate how long it takes for the system to reach the axion-supported scaling solution if the fraction of the energy density in the axion kinetic energy is initially small but non-zero. We first consider initial conditions at time $t_i$ in which the energy density is dominated by the saxion kinetic energy, with $\dot\phi>0$, and denote by
\begin{equation}
\epsilon_V=\frac{V}{\rho}\bigg|_{t_i}\, ,\qquad \epsilon_\theta=\frac{\rho_\theta}{\rho}\bigg|_{t_i}=\frac{J^2e^{2\gamma\phi_i}}{2f_0^2a_i^{2(d-1)}\rho_i}\, ,
\label{eq:initial-fractions}
\end{equation}
the initial fractions of the energy density in the potential and in the axion kinetic energy, with $\epsilon_V,\epsilon_\theta\ll1$. Initial conditions with $\dot\phi<0$, or in which the potential or the axion kinetic energy dominates, are discussed below. Depending on $\alpha$, $\gamma$, and the initial fractions, the system reaches the axion-supported scaling cosmology through different sequences of transitions between four solutions: (i) kination in the saxion direction, (ii) kination in the saxion and axion directions, (iii) a tracker solution in which the non-negligible terms are the saxion kinetic energy and the scalar potential, and (iv) the axion-supported scaling cosmology of Sec.~\ref{sec:scaling-solutions}. We first describe solutions (i)--(iii), summarized together with (iv) in Table~\ref{tab:solutions}, and then derive an upper bound on the time it takes the system to reach the axion-supported scaling solution.

\paragraph{(i) Saxion kination.} As long as the potential and the axion kinetic energy are negligible, the evolution is described by the kination solution in the saxion direction,
\begin{equation}
a(t)=a_i\Bigl(\frac{t}{t_i}\Bigr)^{\frac{1}{d-1}}\, ,\qquad \phi(t)=\phi_i\pm\sqrt{\frac{d-2}{d-1}}\,\log\Bigl(\frac{t}{t_i}\Bigr)=\phi_i\pm\sqrt{(d-1)(d-2)}\,\log\Bigl(\frac{a}{a_i}\Bigr)\, ,
\label{eq:kination}
\end{equation}
which has $w=1$ and $\rho\propto a^{-2(d-1)}$. The upper sign describes a saxion moving towards the cusp and the lower sign a saxion moving away from it. Along this solution, the fractions of the energy density in the potential and in the axion kinetic energy evolve as
\begin{equation}
\frac{V}{\rho}\propto\Bigl(\frac{a}{a_i}\Bigr)^{2(d-1)\mp\alpha\sqrt{(d-1)(d-2)}}\, ,\qquad \frac{\rho_\theta}{\rho}\propto\Bigl(\frac{a}{a_i}\Bigr)^{\pm2\gamma\sqrt{(d-1)(d-2)}}\, .
\label{eq:kination-growth}
\end{equation}
We denote the corresponding rates by
\begin{equation}
\lambda_V=2(d-1)-\alpha\sqrt{(d-1)(d-2)}\, ,\quad \mu_V=2(d-1)+\alpha\sqrt{(d-1)(d-2)}\, ,\quad \lambda_K=2\gamma\sqrt{(d-1)(d-2)}\, .
\label{eq:kination-rates}
\end{equation}
Thus, when the saxion moves towards the cusp, the axion kinetic energy always grows relative to the saxion kinetic energy, at the rate $\lambda_K$, while the potential grows if and only if $\lambda_V>0$, i.e.
\begin{equation}
(d-2)\alpha^2<4(d-1)\, .
\label{eq:radial-existence}
\end{equation}
When the saxion moves away from the cusp, the potential always grows, at the rate $\mu_V$, and the axion kinetic energy decays. Kination is therefore never the final state of the evolution when $J\neq0$.

\paragraph{(ii) Saxion-axion kination.} If the axion kinetic energy becomes important while the potential is still negligible, the energy density remains purely kinetic, so that $\rho\propto a^{-2(d-1)}$, $w=1$, and $a\propto t^{1/(d-1)}$ as in Eq.~\eqref{eq:kination}. The scalar fields then move along a geodesic of the cusp, which can be found in closed form,
\begin{equation}
\phi=\phi_b-\frac{1}{\gamma}\log\cosh u\, ,\qquad \theta=\theta_b+\frac{e^{\gamma\phi_b}}{\gamma f_0}\tanh u\, ,\qquad u\equiv\frac{\lambda_K}{2}\log\Bigl(\frac{a}{a_b}\Bigr)=\gamma\sqrt{\frac{d-2}{d-1}}\,\log\Bigl(\frac{t}{t_b}\Bigr)\, ,
\label{eq:axion-saxion-kination}
\end{equation}
where we took $J>0$ without loss of generality, and the energy fractions evolve as
\begin{equation}
\Omega_\theta=\frac{1}{\cosh^2u}\, ,\qquad \Omega_{\dot\phi}=\tanh^2u\, .
\label{eq:axion-saxion-fractions}
\end{equation}
For $a\ll a_b$ this reproduces the incoming kination solution (i) with $\rho_\theta/\rho\simeq4(a/a_b)^{\lambda_K}$. At $a=a_b$ the axion kinetic energy saturates the energy density, and for $a\gg a_b$ the saxion moves away from the cusp with $\rho_\theta/\rho\simeq4(a/a_b)^{-\lambda_K}$. In other words, the centrifugal barrier in Eq.~\eqref{eq:Veff} reflects the saxion at its maximal value $\phi_b$, while the axion moves by the finite amount $2e^{\gamma\phi_b}/(\gamma f_0)$. Starting from the initial conditions \eqref{eq:initial-fractions}, the reflection happens after $N_b\simeq\lambda_K^{-1}\log(4/\epsilon_\theta)$ $e$-folds at $\phi_b\simeq\phi_i+(2\gamma)^{-1}\log(1/\epsilon_\theta)$. This solution requires $V/\rho\ll1$, and since $V/\rho$ grows as $a^{\mu_V}$ after the reflection, it always ends with the potential becoming important.

\paragraph{(iii) Tracker solution.} If the potential becomes important while the axion kinetic energy is still negligible, and the condition \eqref{eq:radial-existence} holds, the evolution is described by the usual scalar-dominated scaling solution for the potential, which we refer to as the tracker solution \cite{Halliwell:1986ja,Copeland:1997et},
\begin{equation}
a(t)=a_0\Bigl(\frac{t}{t_0}\Bigr)^{\frac{4}{(d-2)\alpha^2}}\, ,\qquad \phi(t)=\frac{1}{\alpha}\log\Bigl(\frac{V_0t^2}{y_V}\Bigr)\, ,\qquad y_V=\frac{2\bigl(4(d-1)-(d-2)\alpha^2\bigr)}{(d-2)\alpha^4}\, .
\label{eq:radial-scaling}
\end{equation}
On this solution $\Omega_{\dot\phi}=\Omega^{\rm tr}_{\dot\phi}\equiv(d-2)\alpha^2/(4(d-1))$, $\Omega_V=1-\Omega^{\rm tr}_{\dot\phi}$, and $w=2\Omega^{\rm tr}_{\dot\phi}-1$. In four dimensions it reduces to $a\propto t^{2/\alpha^2}$, which exists for $\alpha^2<6$. For $J=0$ the tracker solution is the late-time attractor whenever it exists. For $J\neq0$, the ratio of axion kinetic energy to potential energy evolves along it as
\begin{equation}
{\cal R}(a)\equiv\frac{\rho_\theta}{V}={\cal R}_0\Bigl(\frac{a}{a_0}\Bigr)^{\lambda_R}\, ,\qquad \lambda_R=\frac{(d-2)\alpha^2(1+\beta)-4(d-1)}{2}\, .
\label{eq:R-growth}
\end{equation}
Therefore, whenever the existence condition \eqref{eq:rapidturn-existence-d} holds, any non-zero axion momentum grows relative to the tracker solution and eventually becomes important. Hence, the axion-supported branch exists precisely when the tracker solution is unstable to non-zero axion momentum, and the system is eventually driven towards the axion-supported scaling attractor, in agreement with Refs.~\cite{Christodoulidis:2019jsx,Revello:2023hro}. While ${\cal R}\ll{\cal R}_{\rm AS}$, the axion kinetic energy is a spectator, and the equation of state remains that of the tracker solution, $w=2\Omega^{\rm tr}_{\dot\phi}-1$, even though ${\cal R}$ grows. As the trajectory approaches the axion-supported scaling solution, the energy fractions approach their constant attractor values, with ${\cal R}\rightarrow{\cal R}_{\rm AS}$ and $w\rightarrow(1-\beta)/(1+\beta)$. Both values of $w$ follow from the general relation $w=1-2\Omega_V$, which holds for any combination of saxion and axion kinetic energy.

\begin{table}[t]
\centering
\small
\begin{tabular}{lcccc}
\hline\hline
Solution & $a(t)\propto$ & $\phi(t)$ & $(\Omega_{\dot\phi},\Omega_V,\Omega_\theta)$ & Condition \\
\hline
(i) Saxion kination & $t^{\frac{1}{d-1}}$ & $\pm\sqrt{\frac{d-2}{d-1}}\log t$ & $(1,0,0)$ & $\Omega_V,\Omega_\theta\ll1$ \\
(ii) Saxion-axion kination & $t^{\frac{1}{d-1}}$ & $\phi_b-\frac{1}{\gamma}\log\cosh u$ & $(\tanh^2u,0,\cosh^{-2}u)$ & $\Omega_V\ll1$ \\
(iii) Tracker & $t^{\frac{4}{(d-2)\alpha^2}}$ & $\frac{2}{\alpha}\log t$ & $(\Omega^{\rm tr}_{\dot\phi},1-\Omega^{\rm tr}_{\dot\phi},0)$ & $(d-2)\alpha^2<4(d-1)$ \\
(iv) Axion-supported & $t^{\frac{1+\beta}{d-1}}$ & $\frac{2}{\alpha}\log t$ & $(\Omega^{\rm AS}_{\dot\phi},\frac{\beta}{1+\beta},\frac{1}{1+\beta}-\Omega^{\rm AS}_{\dot\phi})$ & $(d-2)\alpha^2(1+\beta)>4(d-1)$ \\
\hline\hline
\end{tabular}
\caption{The four solutions of the axion-saxion system at the cusp. Here $\phi(t)$ is given up to additive constants, $u=(\lambda_K/2)\log(a/a_b)$, $\Omega^{\rm tr}_{\dot\phi}=(d-2)\alpha^2/(4(d-1))$, and $\Omega^{\rm AS}_{\dot\phi}=4(d-1)/((d-2)\alpha^2(1+\beta)^2)$. Solution (i) is unstable to the axion kinetic energy when $\dot\phi>0$ and to the potential when $\dot\phi<0$, solution (ii) always ends when the potential becomes important, solution (iii) is unstable to the axion kinetic energy whenever (iv) exists, and (iv) is the late-time attractor.}
\label{tab:solutions}
\end{table}

\paragraph{Transition time.} We now estimate the number of $e$-folds $N_{\rm AS}$ before the onset of the axion-supported regime, which we take to occur when the system enters a neighborhood of the axion-supported scaling solution, i.e. when the energy fractions and the saxion velocity $\dot\phi/H$ become of the order of their values on this solution. Since the scaling solution is only approached asymptotically, the precise choice of this neighborhood only affects subleading corrections to $N_{\rm AS}$. Note that the ratio ${\cal R}$ alone does not identify this regime, since it can reach its value \eqref{eq:RRT} on the axion-supported scaling solution during a kination phase, while both the potential and the axion kinetic energy are still negligible. On the tracker solution, however, $\Omega_V$ and $\Omega_{\dot\phi}$ are already of order one, and the onset occurs when ${\cal R}$ reaches ${\cal R}_{\rm AS}$. For the initial conditions above, there are three distinct routes.

(a) \textit{Kination $\rightarrow$ tracker $\rightarrow$ axion-supported.} If $\lambda_V>0$ and the potential becomes important before the axion kinetic energy, i.e.
\begin{equation}
\epsilon_\theta\ll\epsilon_V^{\lambda_K/\lambda_V}\, ,
\label{eq:sequence-conditions}
\end{equation}
the kination phase ends after $N_K\simeq\lambda_V^{-1}\log(1/\epsilon_V)$ $e$-folds. Since ${\cal R}\propto a^{\lambda_K-\lambda_V}$ during kination, at this point ${\cal R}_1\simeq\epsilon_\theta\,\epsilon_V^{-\lambda_K/\lambda_V}$, and the tracker phase lasts $N_{\rm tr}\simeq\lambda_R^{-1}\log({\cal R}_{\rm AS}/{\cal R}_1)$ $e$-folds. In total,
\begin{equation}
N_{\rm AS}^{(a)}\simeq\frac{1}{\lambda_V}\log\Bigl(\frac{1}{\epsilon_V}\Bigr)+\frac{1}{\lambda_R}\log\Bigl(\frac{{\cal R}_{\rm AS}\,\epsilon_V^{\lambda_K/\lambda_V}}{\epsilon_\theta}\Bigr)\, .
\label{eq:Nrapidturn}
\end{equation}

(b) \textit{Kination $\rightarrow$ saxion-axion kination $\rightarrow$ tracker $\rightarrow$ axion-supported.} If $\lambda_V>0$ but $\epsilon_\theta\gg\epsilon_V^{\lambda_K/\lambda_V}$, the axion kinetic energy becomes important first, and the saxion is reflected by the centrifugal barrier after $N_b\simeq\lambda_K^{-1}\log(4/\epsilon_\theta)$ $e$-folds, when $V/\rho\simeq\epsilon_V e^{\lambda_VN_b}$. The saxion then moves away from the cusp, and the potential becomes important after $N_{\rm out}\simeq\mu_V^{-1}\log\bigl(e^{-\lambda_VN_b}/\epsilon_V\bigr)$ further $e$-folds. During this phase $\rho_\theta/\rho$ decreases to $\simeq4e^{-\lambda_KN_{\rm out}}$. The potential then turns the saxion around, and the system approaches the tracker solution with ${\cal R}\simeq4e^{-\lambda_KN_{\rm out}}$, from which it takes $\lambda_R^{-1}\log\bigl({\cal R}_{\rm AS}e^{\lambda_KN_{\rm out}}/4\bigr)$ $e$-folds to reach the axion-supported regime. In total,
\begin{equation}
N_{\rm AS}^{(b)}\simeq N_b+\Bigl(1+\frac{\lambda_K}{\lambda_R}\Bigr)N_{\rm out}+\frac{1}{\lambda_R}\log\Bigl(\frac{{\cal R}_{\rm AS}}{4}\Bigr)\, .
\label{eq:Nrapidturn-b}
\end{equation}

(c) \textit{Steep potentials.} If $(d-2)\alpha^2>4(d-1)$, there is no tracker solution and $V/\rho$ decreases during the incoming kination phase. The saxion is reflected by the centrifugal barrier after $N_b\simeq\lambda_K^{-1}\log(4/\epsilon_\theta)$ $e$-folds, and then moves away from the cusp for $N_{\rm out}\simeq\mu_V^{-1}\log\bigl(e^{|\lambda_V|N_b}/\epsilon_V\bigr)$ $e$-folds, until the potential turns it around. The saxion then returns towards the cusp in a kination phase. Since $\rho_\theta/\rho\propto e^{2\gamma\phi}$ during kination, it reaches the centrifugal barrier again after roughly $N_{\rm out}$ $e$-folds, by which time $V/\rho$ has decreased by $e^{-|\lambda_V|N_{\rm out}}$. Each subsequent excursion is therefore shorter by a factor $|\lambda_V|/\mu_V<1$, and the system undergoes damped oscillations around the axion-supported scaling solution, on which $\phi$ continues to grow logarithmically, see Eq.~\eqref{eq:RT-solution}, and converges to it. Summing the geometric series and using $\mu_V-|\lambda_V|=4(d-1)$ and $2(d-1)+|\lambda_V|=\alpha\sqrt{(d-1)(d-2)}$, we find
\begin{equation}
N_{\rm AS}^{(c)}\simeq N_b+\frac{2N_{\rm out}}{1-|\lambda_V|/\mu_V}\simeq\frac{1}{2(d-1)}\Bigl[\log\Bigl(\frac{1}{\epsilon_V}\Bigr)+\frac{1}{\beta}\log\Bigl(\frac{4}{\epsilon_\theta}\Bigr)\Bigr]\, .
\label{eq:Nrapidturn-c}
\end{equation}

(d) \textit{Other initial conditions.} If the saxion initially moves away from the cusp, $\dot\phi<0$, the potential grows at the rate $\mu_V$ and becomes important after $N_{\rm out}\simeq\mu_V^{-1}\log(1/\epsilon_V)$ $e$-folds, while $\rho_\theta/\rho$ decreases to $\simeq\epsilon_\theta e^{-\lambda_KN_{\rm out}}$. This is the second part of route (b), with $N_b=0$ and the factor $4$ replaced by $\epsilon_\theta$. If the tracker solution exists, the system then approaches it with ${\cal R}\simeq\epsilon_\theta e^{-\lambda_KN_{\rm out}}$, and in total
\begin{equation}
N_{\rm AS}^{(d)}\simeq\frac{1}{\lambda_R}\log\Bigl(\frac{{\cal R}_{\rm AS}}{\epsilon_\theta}\Bigr)+\frac{\lambda_R+\lambda_K}{\lambda_R\,\mu_V}\log\Bigl(\frac{1}{\epsilon_V}\Bigr)\, .
\label{eq:Nrapidturn-d}
\end{equation}
For steep potentials, the saxion instead returns to its initial value after another $N_{\rm out}$ $e$-folds, by which time $V/\rho\simeq\epsilon_V^{|\lambda_V|/\mu_V}$ and $\rho_\theta/\rho\simeq\epsilon_\theta$, and then follows route (c). Using $\mu_V-|\lambda_V|=4(d-1)$, the total number of $e$-folds is again given by Eq.~\eqref{eq:Nrapidturn-c}. If the axion kinetic energy initially dominates, the saxion is close to its reflection point, and the evolution follows route (b) or (c) with $N_b\simeq0$. If instead the potential initially dominates, it accelerates the saxion towards the cusp, and after a number of $e$-folds that does not depend on the initial fractions, the system reaches the tracker solution with ${\cal R}\simeq\epsilon_\theta$ or, for steep potentials, a kination phase with $\dot\phi>0$. The subsequent evolution then follows route (a) or (c), respectively, with $\epsilon_V$ of order one.

In Eqs.~\eqref{eq:Nrapidturn}--\eqref{eq:Nrapidturn-d} we assumed the hierarchies that define each route, and a stage that is already completed at $t_i$ has zero duration, i.e. the corresponding logarithm is replaced by zero when its argument is smaller than one. We can now find an upper bound on $N_{\rm AS}$ over all routes and initial conditions. Whenever the tracker solution exists, one has
\begin{equation}
\lambda_K-\lambda_R=\Bigl(2\sqrt{d-1}-\sqrt{d-2}\,\alpha\Bigr)\Bigl(\sqrt{d-2}\,\gamma+\sqrt{d-1}+\frac{\sqrt{d-2}}{2}\alpha\Bigr)>0\, .
\end{equation}
Consequently, the coefficient of $\log(1/\epsilon_V)$ in Eq.~\eqref{eq:Nrapidturn}, $\lambda_V^{-1}(1-\lambda_K/\lambda_R)$, is negative, and the coefficient of $\log(1/\epsilon_\theta)$ in Eq.~\eqref{eq:Nrapidturn-b} is smaller than $\lambda_R^{-1}$. Both routes (a) and (b) are therefore bounded by the right-hand side of Eq.~\eqref{eq:Nrapidturn-d}, and combining with route (c) and the other initial conditions we obtain
\begin{equation}
N_{\rm AS}\lesssim\begin{cases}\displaystyle\frac{1}{\lambda_R}\log\Bigl(\frac{{\cal R}_{\rm AS}}{\epsilon_\theta}\Bigr)+\frac{\lambda_R+\lambda_K}{\lambda_R\,\mu_V}\log\Bigl(\frac{1}{\epsilon_V}\Bigr)\, , & (d-2)\alpha^2<4(d-1)\, ,\\[3mm] \displaystyle\frac{1}{2(d-1)}\Bigl[\log\Bigl(\frac{1}{\epsilon_V}\Bigr)+\frac{1}{\beta}\log\Bigl(\frac{1}{\epsilon_\theta}\Bigr)\Bigr]\, , & (d-2)\alpha^2>4(d-1)\, ,\end{cases}
\label{eq:NRT-bound}
\end{equation}
where these expressions retain the leading logarithmic dependence on the initial energy fractions. At the marginal value $(d-2)\alpha^2=4(d-1)$, the leading logarithmic terms of the two expressions coincide, so that either can be used. For convenience, we recall that the quantities appearing in Eq.~\eqref{eq:NRT-bound} are given in terms of $d$, $\alpha$, and $\gamma$ by
\begin{align}
{\cal R}_{\rm AS}&=\frac{(d-2)\alpha(\alpha+2\gamma)-4(d-1)}{2(d-2)\gamma(\alpha+2\gamma)}\, ,\qquad \mu_V=2(d-1)+\alpha\sqrt{(d-1)(d-2)}\, ,\qquad \beta=\frac{2\gamma}{\alpha}\,, \nonumber\\ \lambda_R&=\frac{(d-2)\alpha(\alpha+2\gamma)-4(d-1)}{2}\, ,\qquad \lambda_K=2\gamma\sqrt{(d-1)(d-2)}\, .
\label{eq:NRT-bound-coefficients}
\end{align}
For $(d-2)\alpha^2<4(d-1)$, the bound is saturated by initial conditions with $\dot\phi<0$, see Eq.~\eqref{eq:Nrapidturn-d}. In addition, the dependence on $\epsilon_\theta$ is saturated when the system starts close to the tracker solution, and the dependence on $\epsilon_V$ by route (b). In particular, the transition time depends only logarithmically on the initial fractions, but for fixed $\epsilon_\theta$ it grows without bound as $\epsilon_V\rightarrow0$, since the saxion, after its reflection or if it initially moves away from the cusp, travels far away from the cusp before the potential can turn it around. Here we assume that the entire excursion remains within the asymptotic region in which Eq.~\eqref{eq:general-rapidturn} holds. The corresponding cosmic time follows from $t\propto a^{d-1}$ during the kination phases (i) and (ii) and $t\propto a^{(d-2)\alpha^2/4}$ on the tracker solution,
\begin{equation}
\frac{t_{\rm AS}}{t_i}\simeq\exp\Bigl[(d-1)N_{\rm kin}+\frac{(d-2)\alpha^2}{4}N_{\rm tr}\Bigr]\, ,
\label{eq:trapidturn}
\end{equation}
where $N_{\rm kin}$ and $N_{\rm tr}$ are the numbers of $e$-folds spent in kination and on the tracker solution, respectively.

Classically, one could take the conserved angular momentum $J$, and hence $\epsilon_\theta$, arbitrarily small and postpone this transition indefinitely. As we will see in Sec.~\ref{sec:localized}, this is no longer the case quantum mechanically. For a localized axion, the root-mean-square axionic momentum is bounded from below, and hence, for a given initial fraction $\epsilon_V$, there is a maximum delay.

\section{Quantum effects at the cusp}
\label{sec:quantum}

In this section we couple the moduli space quantum mechanics of Refs.~\cite{Anchordoqui:2025izb,Anchordoqui:2026nit} to gravity, and show that the finite volume of space and the geometry of the cusp make quantum effects unavoidable for semiclassical cosmological states.

\subsection{The Wheeler-DeWitt equation}
\label{sec:WDW}

Our starting point is the low-energy effective field theory of scalar fields $\phi^i$ (moduli) coupled to gravity,\footnote{Note that in~\cite{Anchordoqui:2025izb,Anchordoqui:2026nit} the scalar fields were commonly denoted by $t^i$.}
\begin{equation}
S_{\rm EFT}\sim\int d^dx \sqrt{-h}\biggl( M_p^{d-2}R\lbrack h_{\mu\nu}\rbrack - {1\over 2}g_{ij}(\phi^i)h^{\mu\nu}\partial_\mu \phi^i\partial_\nu \phi^j-V(\phi^i)\biggr)\, ,
\label{eft}
\end{equation}
where $h_{\mu\nu}$ is the spacetime metric, $R\lbrack h_{\mu\nu}\rbrack$ is the spacetime Ricci scalar, the scalar fields $\phi^i$ correspond to the moduli of the string compactification, for example the geometric moduli of a six-dimensional compact space, $g_{ij}(\phi^i)$ is the moduli space metric, and $V(\phi^i)$ is a possible moduli-dependent scalar potential. Note that by moduli we refer to all light scalar fields in the low-energy effective field theory. The effective field theory could also contain higher-spin fields, but for simplicity we restrict our focus to the scalar fields.

From this action we derive the canonical momenta of the scalar fields,
\begin{equation}
\pi_{\phi^i}={\delta {\cal L}_{\rm EFT}\over \delta(\partial_{t} \phi^i)}= -\sqrt{-h}\,h^{0\mu}g_{ij} \partial_\mu \phi^j\, .
\label{momentum}
\end{equation}
Assuming a foliation with $h^{00}=-1$ and $h^{0\mu}=0$ for $\mu\neq0$, the canonical momenta become
\begin{equation}
\pi_{\phi^i}={\delta {\cal L}_{\rm EFT}\over \delta(\partial_{t} \phi^i)}= \sqrt{-h}\,g_{ij} \dot \phi^j\, ,
\label{momentum1}
\end{equation}
where dots refer to time derivatives. Note that, in contrast to flat spacetime, the canonical momentum contains the spacetime volume factor $\sqrt{-h}$.

In the following we consider a $d$-dimensional isotropic FLRW metric,
\begin{equation}
ds^2=-dt^2+\bigl( a(t)\bigr)^2d\Omega^2_k\, ,
\label{FLRWmetric}
\end{equation}
with $a(t)$ the scale factor and $d\Omega^2_k$ the metric of an isotropic, homogeneous $(d-1)$-dimensional space with constant curvature $k$. For $k=0$ and $d=4$, it can for example be the metric of a three-dimensional torus $T^3$. In cosmic time $t$, and with the volume element of $d\Omega^2_k$ absorbed into the spatial measure, the action for the scalar fields, with the spatial-gradient terms omitted since they do not affect the canonical momenta, then reads
\begin{equation}
S_{\rm scalar}\sim \int d^dx \bigl(a(t)\bigr)^{d-1}\biggl(  g_{ij}(\phi^i){\dot \phi^i\dot \phi^j\over 2}-V(\phi^i)\biggr)\, ,
\label{action2}
\end{equation}
and the canonical momenta become
\begin{equation}
\pi_{\phi^i}={\delta {\cal L}_{\rm EFT}\over \delta(\partial_{t} \phi^i)}= \bigl(a(t)\bigr)^{d-1}g_{ij} \dot \phi^j\, .
\label{momentum2}
\end{equation}

\subsubsection{Quantum commutators}
\label{sec:commutators}

With the action \eqref{eft} and the canonical momenta \eqref{momentum1}, one can canonically quantize the fields $\phi^i$ and their conjugate momenta $\pi_{\phi^i}$~\cite{Anchordoqui:2025izb}
(for more details about the quantization in an FLRW background, see Appendix~\ref{sec:appA}). Concretely, we impose the canonical equal-time commutation relation on the pair $(\phi^i,\pi_{\phi^i})$ of conjugate variables. Since we are now considering QFT in a $d$-dimensional curved spacetime, the commutator involves the $d-1$ spatial coordinates,
\begin{equation}
\lbrack \phi^i(\vec x),\pi_{\phi^j}(\vec y)\rbrack =\lbrack \phi^i(\vec x), \sqrt{-h}\,g_{jk}\dot \phi^k(\vec y)\rbrack =i  \delta^i_j\delta^{(d-1)}(\vec x -\vec y)\, .
\end{equation}
For example, for a $(1+3)$-dimensional FLRW universe we get
\begin{equation}
\lbrack \phi^i(\vec x),\pi_{\phi^j}(\vec y)\rbrack =\lbrack \phi^i(\vec x), (a(t))^3g_{jk}\dot \phi^k(\vec y)\rbrack =i  \delta^i_j\delta^{(3)}(\vec x -\vec y)\, ,
\end{equation}
which can simply be rewritten as
\begin{equation}
\lbrack \phi^i(\vec x), g_{jk}\dot \phi^k(\vec y)\rbrack =i  (a(t))^{-3}\delta^i_j\delta^{(3)}(\vec x -\vec y)\, .
\label{commutatora3}
\end{equation}
Since $(a(t))^{-3}\delta^{(3)}(\vec x-\vec y)$ is the delta function with respect to the physical spatial volume, the commutator of fields averaged over a fixed comoving region is suppressed by the inverse of its physical volume and vanishes in the limit $a\rightarrow\infty$. This applies in particular to the homogeneous modes, see Eq.~\eqref{eq:zero-mode-commutator}.

\subsubsection{The Wheeler-DeWitt equation}
\label{sec:WDWeq}

For the moduli space quantum mechanics of the scalar fields $\phi^i$ alone, we considered the following Schr\"odinger equation~\cite{Anchordoqui:2025izb,Anchordoqui:2026nit}:
\begin{equation}
{\cal H}_{\rm matt}(\phi^i) \psi_{\rm matt}(\phi^i)=E_{\rm matt} \psi_{\rm matt}(\phi^i)\, ,
\label{schrodinger}
\end{equation}
where the wave functions $\psi_{\rm matt}(\phi^i)$ are the eigenfunctions of the matter Hamiltonian ${\cal H}_{\rm matt}$ and $E_{\rm matt}$ are the corresponding energy eigenvalues. The matter Hamiltonian is given by the sum of the kinetic energy, i.e. minus one half of the Laplacian $\Delta_{{\cal M}_{\phi^i}}$ on moduli space, and the potential energy,
\begin{equation}
{\cal H}_{\rm matt}(\phi^i)=-{1\over2}\Delta_{{\cal M}_{\phi^i}}+V(\phi^i)\, .
\end{equation}
The factor $1/2$ corresponds to the canonically normalized kinetic term in Eq.~\eqref{action2}. In quantum gravity, the metric is also treated as a dynamical quantum degree of freedom. Hence, the wave function depends on the scalar fields $\phi^i$ and also on the spacetime metric, $\Psi=\Psi(h_{\mu\nu},\phi^i)$, which in the cosmological setting reduces to a dependence on the scale factor $a$. We employ the ADM (Arnowitt-Deser-Misner) formalism \cite{Misner:1967uu} and, instead of the Schr\"odinger equation, we now consider the Wheeler-DeWitt (WDW) equation~\cite{Wheeler:1968iap,DeWitt:1967yk}. In the presence of matter fields, namely the scalar fields $\phi^i$ with moduli space metric $g_{ij}(\phi^i)$ and potential $V(\phi^i)$, one schematically obtains
\begin{equation}
\Bigl( -G_{abcd}{\delta^2\over\delta h_{ab}\delta h_{cd}}-{1\over \sqrt h}g^{ij}(\phi^i){\delta^2\over\delta \phi^i\delta \phi^j}-\sqrt h\, {R}^{(d-1)}+\sqrt h\, V(t)\Bigr)\Psi(h,\phi^i)=0\, ,
\label{wdewitt}
\end{equation}
where $h_{ab}$ is the spatial metric, $G_{abcd}$ is the DeWitt supermetric, and ${R}^{(d-1)}$ is the spatial Ricci scalar. In an isotropic universe with scale factor $a$ this takes the schematic form
\begin{equation}
{\cal H}_{\rm total}(a,\phi^i) \Psi(a,\phi^i)=\bigl({\cal H}_{\rm grav}(a) +{\cal H}_{\rm matt}(a,\phi^i)\bigr)\Psi(a,\phi^i)=0\, .
\label{wdewitta}
\end{equation}
Some remarks are in order to justify this truncation and to clarify its physical meaning. We can dimensionally reduce FLRW spacetimes to $0+1$ dimensions. For spatially flat spacetimes, we can take the comoving coordinates to be periodic with period one and interpret the spatial slices as an isotropic torus $T^{d-1}$ with volume $a^{d-1}$ in $d$-dimensional Planck units. Similarly, for spacetimes with spatial curvature parameter $k=-1$, we can take the comoving coordinates to parameterize a locally isotropic compact manifold, namely a compact quotient of hyperbolic space. For $k=+1$ the spatial slice is naturally compact. More generally, we denote by $V_c$ the comoving volume of the spatial slice, so that its physical volume is $V_ca^{d-1}$, with $V_c=1$ for the tori above.

In all these cases, Eq.~\eqref{wdewitta} is nothing other than the Wheeler-DeWitt equation of a one-dimensional quantum gravity theory restricted to a section of its moduli space, specifically the higher-dimensional moduli together with the overall volume modulus. In other words, we assume that the wave packet is peaked around an isotropic configuration, which is why we do not include the shape moduli, and we postpone a more complete treatment of the other moduli and of the inhomogeneous modes to future work. This provides an approximate $0+1$-dimensional interpretation of the wave function, obtained by restricting the full higher-dimensional quantum-gravity wave functional to the homogeneous and isotropic minisuperspace sector.

Corrections from shape moduli and inhomogeneous modes are neglected, which requires additional assumptions on the initial state and on the dynamics, since, for example, the homogeneous shear energy scales as $a^{-2(d-1)}$, like the kinetic energy of the moduli. Finite-volume quantum effects, such as the Casimir energy, are also neglected and become important only when $a$ becomes sufficiently small. Note that if the higher-dimensional theory contains a Kalb-Ramond $B$-field, dimensional reduction of the metric and the two-form gauge field enlarges the torus moduli space to the $O(d-1,d-1)$ Narain moduli space, whose Wheeler-DeWitt equation was studied in Ref.~\cite{Gasperini:1996fu}. As we will see, including the volume modulus already captures the gravitational effects relevant for cosmology, such as the backreaction of the scalar sector.

It has been argued that, for compact spatial slices, the exact Hilbert space of quantum gravity is one-dimensional~\cite{McNamara:2020uza}. However, an internal observer can have a non-trivial effective Hilbert-space description, while there can be no non-trivial external observer within this framework \cite{Harlow:2025pvj}. Identifying an observer inside the closed universe may require substantially more structure than our restriction of the wave function to specific configurations, such as homogeneous ones. Nevertheless, the underlying principle stands.  Establishing the precise relation between our minisuperspace description and such an internal observer's effective Hilbert space is outside the scope of this work.\footnote{We thank Jacob McNamara for explaining this point to us.}

Here the matter Hamiltonian ${\cal H}_{\rm matt}(a,\phi^i)$ acts as an energy source, and ${\cal H}_{\rm matt}(a,\phi^i)/(V_ca^{d-1})$ corresponds in classical gravity to the homogeneous energy density $T_{00}(a,\phi^i)$ that sources the Friedmann equation. Since the time coordinate in general relativity is arbitrary, the total energy of the universe, i.e. the sum of the gravitational and matter energies, always vanishes, and the gravitational energy can be negative.

\subsubsection{Gravitational part of the Wheeler-DeWitt equation}
\label{sec:WDWgrav}

Let us first recall the minisuperspace approximation for the well-known case of a four-dimensional isotropic universe with a constant potential $V\equiv \Lambda$, where $\Lambda$ is the cosmological constant. Working in reduced Planck units, the momentum conjugate to $a$ is
\begin{equation}
\pi_a=-6V_ca\dot a\, ,
\end{equation}
and the gravitational part of the WDW equation reduces to a one-dimensional equation for the scale factor $a$ (see e.g. \cite{Gielen:2023hsg})
\begin{equation}
\biggl({1\over 12 a}{d^2\over da^2}-{1\over 12a^2}{d\over da}-V_c^2(3ka-a^3\Lambda) \biggr)\psi_{\rm grav}(a)=0\, .
\label{WDW}
\end{equation}
For $k=+1$, the natural normalization is the volume of the unit three-sphere, $V_c=2\pi^2$, whereas for the spatially flat compact slices considered in the rest of this paper we take $V_c=1$. The WDW equation can be rewritten as
\begin{equation}
\biggl(-{d^2\over da^2}+{1\over a}{d\over da}+U(a) \biggr)\psi_{\rm grav}(a)=0\, ,
\label{WDW5}
\end{equation}
with the WDW potential
\begin{equation}
U(a)=36V_c^2(ka^2-H^2a^4)\, ,\qquad H^2={\Lambda\over 3}\, .
\label{WDWpotential}
\end{equation}
In terms of the variable $a^2$, Eq.~\eqref{WDW5} becomes the Airy equation, and its general solution has the form (see e.g. \cite{Gielen:2023hsg})
\begin{equation}
\psi_{\rm grav}(a)=c_1\, {\rm Ai}\biggl({(3V_c^2)^{1/3}(3k-a^2\Lambda)\over\Lambda^{2/3}}\biggr)+c_2\, {\rm Bi}\biggl({(3V_c^2)^{1/3}(3k-a^2\Lambda)\over\Lambda^{2/3}}\biggr)\, ,
\end{equation}
where $c_1$ and $c_2$ are constants and ${\rm Ai}$ and ${\rm Bi}$ are the Airy functions.

De Sitter space admits slicings with $k=-1$, $0$, or $+1$. For $k=+1$, which corresponds to the global slicing of de Sitter space, the potential $U(a)$ vanishes at $a=1/H$. Classically, $a(x^0)=H^{-1}\cosh(H x^0)$, which has the minimum value $1/H$. Here $H$ denotes the Hubble scale of de Sitter space, which coincides with $\dot a/a$ only for $k=0$, whereas for $k=+1$ one has $\dot a/a=H\tanh(Hx^0)$. In the classically forbidden region $a<1/H$, the WKB approximation with $|p(a)|=\sqrt{U(a)}$ gives the general solution
\begin{equation}
\psi_{\rm grav}(a)=\sqrt{a\over|p(a)|}\,\biggl[c_1 \exp\biggl(\int_a^{1/H}|p(a')|da'\biggr)+c_2 \exp\biggl(-\int_a^{1/H}|p(a')|da'\biggr)\biggr]\, .
\end{equation}
In the classically allowed region $a>1/H$, the two solutions are
\begin{equation}
\psi_{\rm grav}(a)=\sqrt{a\over p(a)}\,\biggl[c_1 \exp\biggl(i\int_{1/H}^a p(a')da'\biggr)+c_2 \exp\biggl(-i\int_{1/H}^a p(a')da'\biggr)\biggr]\, ,
\end{equation}
with $p(a)=\sqrt{-U(a)}$. The choice of the two constants depends on the boundary conditions and on the interpretation of the wave function. Vilenkin's wave function \cite{Vilenkin:1982de} contains only the outgoing (expanding) solution in the classically allowed regime. For large scale factors $a$, where $p(a)\simeq 6V_cHa^2$, it takes the form
\begin{equation}
\psi_{\rm grav,V}(a)\simeq {1\over\sqrt a}\exp\bigl(-2iV_c\,Ha^3\bigr)\,.
\end{equation}

On the other hand, the Hartle-Hawking (HH) wave function \cite{Hartle:1983ai} is real and is interpreted as the amplitude for the creation of a universe of size $a$ from nothing. In the classically forbidden region it is the solution that decreases towards small $a$, $\psi_{\rm grav,HH}(a)\propto\sqrt{a/|p(a)|}\,\exp\bigl(-\int_a^{1/H}|p(a')|da'\bigr)$. In the classically allowed region it becomes a superposition of incoming and outgoing waves with equal weights,
\begin{equation}
\psi_{\rm grav,HH}(a)\simeq \Bigl({\Lambda\over 3}a^2-k\Bigr)^{-1/4}\cos\biggl({6V_c\over \Lambda}\Bigl({\Lambda\over 3}a^2-k\Bigr)^{3/2}-{\pi\over 4}\biggr)\,.
\end{equation}
The overall amplitude is determined by the Euclidean action of the de Sitter instanton, i.e. half of the Euclidean four-sphere,
\begin{equation}
{\cal S}_E=- {3\pi \over 2G\Lambda}=-{12\pi^2\over\Lambda}\, ,
\end{equation}
where in the last step we used reduced Planck units, $8\pi G=1$, as in Eq.~\eqref{WDW}. The associated HH amplitude at the nucleation point $a=1/H$ takes the form
\begin{equation}
\psi_{\rm grav, HH}(1/H)\sim e^{-{\cal S}_E}=e^{12\pi^2 \over \Lambda}\, ,
\end{equation}
in agreement with the WKB exponent $\int_0^{1/H}|p(a')|da'=6V_c/\Lambda$ for the unit three-sphere, $V_c=2\pi^2$. Here the sign in the exponential is such that a small cosmological constant, i.e. a very negative instanton action, leads to a large amplitude.

\subsubsection{Comment on the signature of the kinetic term}
\label{sec:signature}

In the minisuperspace WDW equation \eqref{WDW}, and in its extension \eqref{WDW1} with moduli, the kinetic term of the scale factor enters with a sign opposite to that of the moduli kinetic terms. In infinite-distance limits of moduli space it is often observed that the moduli kinetic terms, written in the species frame, acquire a ``wrong-sign'' direction of their own. The string-frame dilaton is the standard example. In this subsection we explain when such a wrong-sign direction appears and why the frame-independent statement concerns the combined gravity-plus-moduli kinetic term rather than the moduli alone.

\paragraph{Species-frame moduli metric.} Consider the action \eqref{eft} with $M_p=1$ and a canonically normalized Einstein-Hilbert term,
\begin{equation}
S=\int d^dx\sqrt{-h}\,\Bigl[\tfrac12R-\tfrac12g_{ij}\,\partial \phi^i\!\cdot\partial \phi^j-V\Bigr]\, ,
\end{equation}
where the moduli space metric $g_{ij}$ is positive definite. Let $\omega=\log\Lambda_s$, where $\Lambda_s$ is the species scale~\cite{Dvali:2007hz,Dvali:2007wp}, and define the species frame by $h^{\rm sp}_{\mu\nu}=e^{2\omega}h_{\mu\nu}$, so that lengths are measured in units of $\Lambda_s^{-1}$. A Weyl rescaling gives
\begin{equation}
S=\int d^dx\sqrt{-h^{\rm sp}}\,N_s\Bigl[\tfrac12R^{\rm sp}-\tfrac12g^{\rm sp}_{ij}\,\partial \phi^i\!\cdot\partial \phi^j-e^{-2\omega}V\Bigr]\, ,
\label{eq:Ssp}
\end{equation}
with the species-frame moduli metric
\begin{equation}
g^{\rm sp}_{ij}=g_{ij}-(d-1)(d-2)\,\partial_i\omega\,\partial_j\omega\, ,
\label{eq:Gsp}
\end{equation}
where $N_s=e^{-(d-2)\omega}$ is the number of species, which plays the role of $e^{-2\Phi}$ in string frame. Since $g^{\rm sp}$ is a rank-one deformation of a positive-definite metric, with
\begin{equation}
\det g^{\rm sp}=\det g\,\bigl(1-(d-1)(d-2)|\nabla\omega|^2\bigr)\, ,\qquad |\nabla\omega|^2\equiv g^{ij}\partial_i\omega\,\partial_j\omega\, ,
\end{equation}
it has at most one non-positive eigenvalue. Hence $g^{\rm sp}$ is Lorentzian, degenerate, or Euclidean if
\begin{equation}
\frac{|\nabla\Lambda_s|}{\Lambda_s}\ >,\ =,\ <\ \frac{1}{\sqrt{(d-1)(d-2)}}\, ,
\label{eq:criterion}
\end{equation}
respectively.

\paragraph{Infinite-distance limits.} Along any infinite-distance limit, the species exponent $\lambda_{\rm sp}=|\nabla\log\Lambda_s|$ obeys \cite{Calderon-Infante:2023ler,Castellano:2023stg,Castellano:2023jjt,vandeHeisteeg:2023ubh}
\begin{equation}
\frac{1}{\sqrt{(d-1)(d-2)}}\leq\lambda_{\rm sp}\leq\frac{1}{\sqrt{d-2}}\, .
\end{equation}
In an emergent-string limit the upper bound is saturated and $(d-1)(d-2)\lambda_{\rm sp}^2=d-1>1$. For the decompactification of $k$ dimensions to $D=d+k$ dimensions, one has $\lambda_{\rm sp}^2=k/[(d-2)(D-2)]$, which gives
\begin{equation}
(d-1)(d-2)\lambda_{\rm sp}^2-1=\frac{(d-2)(k-1)}{D-2}\ \geq\ 0\, .
\end{equation}
The species-frame moduli metric is therefore Lorentzian in every infinite-distance limit, except for the decompactification of a single extra dimension, where it is exactly degenerate. In that case the higher-dimensional frame is Brans-Dicke theory with $\omega_{\rm BD}=0$, and the radion has no kinetic term of its own. The lower bound on the species exponent is thus precisely the statement that the species-frame moduli metric is never Euclidean asymptotically.

\paragraph{Interior of moduli space.} In the interior of moduli space, by contrast, nothing forces $|\nabla\Lambda_s|/\Lambda_s$ to be large. At a desert point, where the species scale is maximized \cite{Long:2021jlv,vandeHeisteeg:2022btw,Cribiori:2022nke,vandeHeisteeg:2023dlw}, 
one has $\nabla\omega=0$. There $g^{\rm sp}=g$ is Euclidean, and $g^{\rm sp}$ changes signature across the hypersurface on which the criterion \eqref{eq:criterion} is saturated. In ten-dimensional type IIB string theory, $\Lambda_s\propto E_{3/2}({\cal T},\bar{\cal T})^{-1/6}$, where ${\cal T}=C_0+ie^{-\Phi}$ is the axio-dilaton and $E_{3/2}$ is the non-holomorphic Eisenstein series multiplying the $R^4$ term \cite{Green:1997tv}. With the canonically normalized axio-dilaton metric one finds $(d-1)(d-2)|\nabla\omega|^2=\bigl(2{\cal T}_2|\partial E_{3/2}|/E_{3/2}\bigr)^2$, where $|\partial E_{3/2}|$ denotes the flat gradient norm in the $({\cal T}_1,{\cal T}_2)$ plane, so the criterion becomes $2{\cal T}_2|\partial E_{3/2}|/E_{3/2}>1$. The left-hand side tends to $3$ at weak coupling, which reproduces the wrong-sign string-frame dilaton, whose metric component is multiplied by $1-9=-8$. It vanishes at the self-dual points ${\cal T}=i$ and ${\cal T}=e^{2\pi i/3}$. Along $C_0=0$ the signature flips at $g_s\simeq0.604$ and at its S-dual value $g_s\simeq1.656$. In between, the full species-frame axio-dilaton metric is positive definite.

\paragraph{Minisuperspace.} A Euclidean $g^{\rm sp}$ does not mean that the theory has lost its timelike direction. The timelike direction has moved into the gravitational sector. This becomes explicit after the reduction to $0+1$ dimensions discussed in Sec.~\ref{sec:WDWeq}. For $ds^2=-N^2dt^2+e^{2\sigma}d\vec x^{\,2}$, with $\sigma=\log a$ and $d\vec x^{\,2}$ the flat metric on a torus of coordinate volume $V_c$, the homogeneous Lagrangian is
\begin{equation}
L=\frac{V_c\,e^{(d-1)\sigma}}{2N}\Bigl[-(d-1)(d-2)\,\dot\sigma^2+g_{ij}\dot \phi^i\dot \phi^j\Bigr]-N\,V_c\,e^{(d-1)\sigma}\,V\, .
\label{eq:Lminisuperspace}
\end{equation}
For $d=4$ and $N=1$, its gravitational part reproduces the momentum $\pi_a=-6V_ca\dot a$ of Sec.~\ref{sec:WDWgrav}. The minisuperspace metric is Lorentzian, with a single timelike direction given by the conformal mode, as for DeWitt's supermetric \cite{DeWitt:1967yk,Gibbons:1978ac}. In $0+1$ dimensions, a Weyl rescaling is no longer a change of conformal frame but simply a field redefinition. The species frame corresponds to $\sigma_{\rm sp}=\sigma+\omega$ and $N_{\rm sp}=e^{\omega}N$, and Eq.~\eqref{eq:Lminisuperspace} becomes
\begin{equation}
L_{\rm kin}=\frac{V_c\,e^{(d-1)\sigma_{\rm sp}}N_s}{2N_{\rm sp}}\Bigl[-(d-1)(d-2)\bigl(\dot\sigma_{\rm sp}-\partial_i\omega\,\dot \phi^i\bigr)^2+g_{ij}\dot \phi^i\dot \phi^j\Bigr]\, .
\label{eq:Lsp}
\end{equation}
The shift of $\sigma$ is a point transformation with a triangular, unit-determinant Jacobian, so it acts on the kinetic matrix by congruence. The rescaling of the lapse, the only component of the one-dimensional metric, multiplies the kinetic matrix by a positive function. By Sylvester's law of inertia, neither operation changes the signature, which remains $(1,n_t)$, i.e. one negative and $n_t$ positive eigenvalues, in every frame, where $n_t$ is the number of moduli. For a single modulus $t$ with metric $g\equiv g_{tt}$ and $\omega'=d\omega/dt$, the kinetic matrix of Eq.~\eqref{eq:Lsp} is proportional to
\begin{equation}
{\cal G}_{\rm sp}=\begin{pmatrix}-A & A\omega'\\ A\omega' & g-A\omega'^2\end{pmatrix}\, ,\qquad A=(d-1)(d-2)\, ,\qquad \det{\cal G}_{\rm sp}=-Ag<0\, .
\end{equation}
The modulus entry $g-A\omega'^2$ can take either sign, but the matrix always has exactly one negative eigenvalue. Geometrically, $g^{\rm sp}$ is the pull-back of the Lorentzian minisuperspace metric to the slices $\sigma_{\rm sp}={\rm const}$ of fixed species-frame volume. Their normal covector $d\sigma+\partial_i\omega\,d\phi^i$ has norm proportional to $-1+(d-1)(d-2)|\nabla\omega|^2$. The slices are therefore timelike, null, or spacelike exactly in the three cases of Eq.~\eqref{eq:criterion}. Einstein-frame slices $\sigma={\rm const}$ are always spacelike, which is why the Einstein-frame moduli metric is always Euclidean. The negative direction rotates between gravity and moduli as one moves through moduli space, but it never disappears. In higher dimensions, the same statement cannot be made about the moduli alone, because the Weyl factor mixes them with the conformal mode of the metric, which is not a finite-dimensional coordinate. This single timelike direction is what makes the WDW equation hyperbolic, with one intrinsic time \cite{DeWitt:1967yk}, and it is the origin of the relative sign between the scale-factor and moduli kinetic terms in Eq.~\eqref{WDW1}. It is also what allows a positive excitation energy of the moduli to act as a positive gravitational source, as in Eq.~\eqref{lambdaeff6} of Sec.~\ref{sec:vanishingV}.

\paragraph{Higher-derivative corrections.} The argument above uses the two-derivative action. With higher-derivative corrections, after the field redefinitions that remove spurious higher time derivatives, the homogeneous Lagrangian takes the form $N\ell(q,v)$, with $q^A=(\sigma,\phi^i)$ and $v^A=\dot q^A/N$. The natural object is then the kinetic Hessian $K_{AB}=\partial^2\ell/\partial v^A\partial v^B$. We expect that $K$ keeps exactly one negative eigenvalue throughout the regime in which the homogeneous description is valid. Below the cutoff this is perturbatively automatic. It becomes a non-trivial statement whenever the homogeneous description is exact. Tree-level string cosmology to all orders in $\alpha'$ is an example. After reduction, the action takes the form \cite{Hohm:2019ccp}
\begin{equation}
I=\int dx^0\,N\,e^{-\bar\Phi}\Bigl[-\bigl(N^{-1}\dot{\bar\Phi}\bigr)^2-F(H)\Bigr]\, ,\qquad F(H)=-(d-1)H^2+{\cal O}(\alpha')\, ,
\end{equation}
where $\bar\Phi$ is the shifted dilaton and $H=N^{-1}\dot a/a$ is the string-frame Hubble rate. The Hessian with respect to the velocities of $(\bar\Phi,\log a)$ is $e^{-\bar\Phi}{\rm diag}\bigl(-2,-F''(H)\bigr)$, so the expectation is $F''(H)<0$ for all $H$. Since the Hamiltonian constraint reads $\dot{\bar\Phi}^2=F(H)-HF'(H)$ for $N=1$, this would exclude every classical change of sign of $\dot{\bar\Phi}$, as well as the non-perturbative string-frame de Sitter solutions of Ref.~\cite{Hohm:2019jgu}, which require $F'(H_0)=0$ at some $H_0\neq0$. In the isotropic sector, the leading corrections have the reinforcing sign \cite{Codina:2021cxh},
\begin{equation}
-F''(H)=\begin{cases} 2(d-1)+6(d-1)\,\alpha'H^2+\dots & \text{(bosonic)}\, ,\\ 2(d-1)+3(d-1)\,\alpha'H^2+\dots & \text{(heterotic)}\, ,\\ 2(d-1)+7(d-1)(2d-5)\,\zeta(3)\,\alpha'^3H^6+\dots & \text{(type II)}\, .\end{cases}
\end{equation}
Determining whether this sign persists beyond the leading order is a sharp test of the expectation above.

\subsection{Warm-up: solutions with vanishing potentials}
\label{sec:vanishingV}

Let us now consider the matter part of the WDW equation, where the matter sector consists of a certain number of scalar fields $\phi^i$. In the presence of a scale factor $a$, the matter Hamiltonian is given by
\begin{equation}
{\cal H}_{\rm matt}(a,\phi^i)=-{1\over 2V_ca^3}\Delta_{{\cal M}_{\phi^i}}+V_ca^3V(\phi^i)\, ,
\label{alapl}
\end{equation}
where $\Delta_{{\cal M}_{\phi^i}}$ is the moduli space Laplacian and $V_ca^3$ is the spatial volume. In $d$ spacetime dimensions the factors $a^{\mp3}$ are replaced by $a^{\mp(d-1)}$. The factor $a^{-3}$ in the kinetic term directly reflects the commutator \eqref{commutatora3}. The cosmological constant $\Lambda$ is now replaced by the scalar-field-dependent potential $V(\phi^i)$, and the total WDW equation becomes
\begin{equation}
V_c\,{\cal H}_{\rm total}(a,\phi^i) \Psi(a,\phi^i)=\biggl({1\over 12 a}{d^2\over da^2}-{1\over 12a^2}{d\over da}-V_c^2\bigl(3ka-a^3V(\phi^i)\bigr)-{1\over 2a^3}\Delta_{{\cal M}_{\phi^i}}\biggr)\Psi(a,\phi^i)=0\, .
\label{WDW1}
\end{equation}
Note that the sign of the kinetic term for the scale factor $a$ is opposite to that of the kinetic terms of the scalar fields $\phi^i$. As explained in Sec.~\ref{sec:signature}, this single negative direction of the combined kinetic term is frame independent.

Let us first focus on the case of vanishing potential, $V(\phi^i)=0$. In this case we make the separation ansatz
\begin{equation}
\Psi(a,\phi^i)=\psi_{\rm grav}(a) \psi_{\rm matt}(a,\phi^i)\, .
\label{Psi}
\end{equation}
We assume that $\psi_{\rm matt}(a,\phi^i)$ is an eigenfunction of ${\cal H}_{\rm matt}(a,\phi^i)=-{1\over 2V_ca^3}\Delta_{{\cal M}_{\phi^i}}$ with eigenvalue $E_{\rm matt}(a)$, i.e.
\begin{equation}
-{1\over 2V_ca^3}\Delta_{{\cal M}_{\phi^i}}\psi_{\rm matt}(a,\phi^i)=E_{\rm matt}(a) \psi_{\rm matt}(a,\phi^i)\, .
\label{schrodinger1}
\end{equation}
From the results of moduli space quantum mechanics~\cite{Anchordoqui:2026nit}, we already know the eigenfunctions and eigenvalues of the Schr\"odinger equation \eqref{schrodinger}. From these we immediately deduce that the energy eigenvalues $E_{\rm matt}(a)$ in the presence of a scale factor $a$ are related to the energy eigenvalues $E_{\rm matt}$ without a scale factor in the following simple way:
\begin{equation}
E_{\rm matt}(a)={E_{\rm matt}\over V_ca^3}\, .
\end{equation}
Thus, for large spatial volumes $V_ca^3$ the energy eigenvalues are suppressed. Concerning the wave function, it is easy to see that the eigenfunctions of the operator $-{1\over 2V_ca^3}\Delta_{{\cal M}_{\phi^i}}$ in Eq.~\eqref{alapl} are the same as the eigenfunctions of $-\Delta_{{\cal M}_{\phi^i}}$ without the scale factor and do not depend on $a$:
\begin{equation}
\psi_{\rm matt}(a,\phi^i)=\psi_{\rm matt}(\phi^i)\, .
\end{equation}

As an example, let us consider a two-dimensional moduli space with two scalar fields $\phi^1$ and $\phi^2$, denoted as $\phi$ and $\theta$, where $\theta$ is a periodic axion and $\phi$ is the corresponding saxion. We are interested in scalar field spaces in which the axion decay constant decreases exponentially with the saxion, i.e. in the presence of a cusp. We consider the metric
\begin{equation}
ds^2 = d\phi^2 +f_0^2 e^{-2\gamma|\phi|}d\theta^2\, ,
\label{2dimmetric}
\end{equation}
where $\phi\in{\mathbb R}$, $\theta \sim \theta +2\pi$, and $\gamma$ is a positive constant that controls the exponential rate at which the radius of the circle shrinks. For $\phi>0$ this is the moduli space metric of the axion-saxion system of Sec.~\ref{sec:scaling}. The moduli space ${\cal M}_{\phi,\theta}$ is invariant under the duality symmetry $\phi\rightarrow-\phi$. Its curvature is $R_{{\cal M}_{\phi,\theta}}=-2\gamma^2$ for $\phi\neq0$, whereas at $\phi=0$ the metric is not smooth, and we assume that it is smoothed out by the interior of the moduli space. In the following we are only interested in the cusp region $\phi>0$. On the $\phi>0$ branch, defining $\varphi=e^{\gamma\phi}$, 
the metric becomes\footnote{The scalar $\varphi$ was denoted by $t$ in~\cite{Anchordoqui:2025izb,Anchordoqui:2026nit}.}
\begin{equation}
ds^2={d\varphi^2\over\gamma^2\varphi^2}+f_0^2{d\theta^2\over \varphi^2}\, .
\end{equation}
Metrics of this type commonly arise in four-dimensional ${\cal N}=2$ compactifications. A special case is $\gamma f_0=1$. In this case, introducing the complex coordinate $\tau=\theta+i\varphi$, the metric becomes ($f_0^2$ times) the standard hyperbolic metric on the upper half-plane. If the theory has the corresponding $SL(2,\mathbb Z)$ duality, the moduli space is
\begin{equation}
{\cal M}_\tau={\mathbb H}^+/SL(2,\mathbb Z)\, ,
\end{equation}
where in this case the axion is normalized to have unit period, $\theta\simeq\theta+1$, which is related to the normalization $\theta\simeq\theta+2\pi$ used elsewhere by $\theta\rightarrow\theta/(2\pi)$ and $f_0\rightarrow2\pi f_0$.

On the $\phi>0$ branch, the moduli space Laplacian in terms of the fields $\phi$ and $\theta$ reads
\begin{equation}
-\Delta_{{\cal M}_{\phi,\theta}} =-\partial^2_\phi+\gamma\partial_\phi-{1\over f_0^2}e^{2\gamma|\phi |}\partial^2_\theta\, .
\label{laplacian}
\end{equation}
In~\cite{Anchordoqui:2026nit} we considered, for $\gamma=f_0=1$, the modular invariant wave functions as eigenfunctions of the modular invariant Laplacian $\nabla^2$ over the two-dimensional fundamental domain of $SL(2,{\mathbb Z})$. These are given in terms of the non-holomorphic Eisenstein series (Maa{\ss} forms) $E_{{1\over 2}+i{\nu\over\sqrt2}}(\tau,\bar\tau)$:
\begin{equation}
\psi_{\nu}(\tau,\bar\tau)=E_{{1\over 2}+i{\nu\over\sqrt2}}(\tau,\bar\tau)=\sqrt{\varphi}\biggl(A_\nu\, \varphi^{i{\nu\over\sqrt2}}+B_\nu\, \varphi^{-i{\nu\over\sqrt2}}+\sum_{n\neq0}c_ne^{2\pi i n\theta}K_{i{\nu\over\sqrt2}}\bigl(2\pi|n|\varphi\bigr)\biggr)\, .
\label{wavefunctionmod}
\end{equation}
The corresponding spectrum of $-\Delta_{{\cal M}_\tau}$ has the form
\begin{equation}
E(\nu)= {1\over 4}+{\nu^2\over 2}\, ,
\label{Ealpha}
\end{equation}
which starts at the threshold $1/4$. One of the interesting properties of these solutions is that the excited spectrum includes infinitely many bound states, which correspond to the square-integrable cusp forms
\begin{equation}
\psi_{\nu_j}^{\rm bound~state}(\tau,\bar\tau)={\cal S}_{{1\over2}+i{\nu_j\over\sqrt2}}(\tau,\bar\tau)=\sqrt{\varphi}\sum_{n\neq0}c_ne^{2\pi i n\theta}K_{i{\nu_j\over\sqrt2}}\bigl(2\pi|n|\varphi\bigr)\, .
\end{equation}
The discrete spectral parameters $\nu_j$ of the bound states are known only numerically \cite{DHoker:2022dxx},
\begin{equation}
\nu_1\simeq 13.48\, ,\quad \nu_2\simeq 17.22\, ,\quad\nu_3\simeq 19.49\, ,\quad \dots
\end{equation}

In the presence of the scale factor $a$, the Hamiltonian is only multiplied by $1/(V_ca^3)$, so that the eigenfunctions, i.e. the constant mode, the cusp forms, and the Eisenstein series, are unchanged. However, the energies are suppressed by the spatial volume and now take the form
\begin{equation}
E(a,\nu)= {1\over 2V_ca^3}\Bigl({1\over 4}+{\nu^2\over 2}\Bigr)\, .
\end{equation}
Hence, for large $a$, the bound states have a smaller energy than in the case without a scale factor.

We now combine $\psi_{\rm grav}(a)$ with the matter wave function $\psi_{\rm matt}(a,\phi^i)=\psi_{\rm matt}(\phi^i)$ as in Eq.~\eqref{Psi}. The WDW equation for the scale factor $a$ and the matter fields $\phi^i$ in the case of vanishing potential takes the form (recall that we have set the bare potential $\Lambda$ to zero)
\begin{equation}
\biggl({1\over 12 a}{d^2\over da^2}-{1\over 12a^2}{d\over da}-3V_c^2ka-{1\over 2a^3}\Delta_{{\cal M}_{\phi^i}}\biggr)\Psi(a,\phi^i)=0\, .
\end{equation}
Using the ansatz \eqref{Psi} together with Eq.~\eqref{schrodinger1}, we immediately obtain
\begin{equation}
\biggl({1\over 12 a}{d^2\over da^2}-{1\over 12a^2}{d\over da}-V_c^2\Bigl(3ka- {E_{\rm matt}\over V_c^2a^3}\Bigr) \biggr)\psi_{\rm grav}(a)=0\, .
\end{equation}
Comparing with Eq.~\eqref{WDW}, we see that the matter energy eigenvalues ${E_{\rm matt}}$ act in the WDW equation as an effective, scale-dependent cosmological constant of size
\begin{equation}
\Lambda_{\rm eff}={E_{\rm matt}\over (V_ca^3)^2}\, ,
\label{lambdaeff6}
\end{equation}
i.e. the matter energy divided by the square of the spatial volume. So the excited matter states, namely the bound states mentioned above, can be regarded as positive energy states with a scale-dependent energy $\Lambda_{\rm eff}$, which is strongly suppressed at large scale factors $a$. Since the scale factor is time dependent, $\Lambda_{\rm eff}$ is also time dependent and disappears at late times. Let us consider the equation of state parameter $w$ for these states. The power $a^{-6}$ corresponds to the scaling of kinetic energy. Since pure kinetic energy corresponds to $w=1$, this does not lead to accelerated expansion, which requires $w<-1/3$. 
We therefore have to include a potential $V(\phi^i)$ and determine how the effective minimum and $w$ depend on the parameters of the potential.

\subsection{Solutions with exponential potentials}
\label{sec:nonvanishingV}

When the potential $V(\phi^i)$ is included, the WDW equation takes the form \eqref{WDW1}. However, due to the presence of the potential, the separation ansatz \eqref{Psi} no longer works. As in the previous case without potential, we consider the two-dimensional moduli space with metric \eqref{2dimmetric}, restricted to the cusp region $\phi>0$. The corresponding Laplacian is given in Eq.~\eqref{laplacian}. In the following we assume that the potential depends only on $\phi$ and does not break the shift symmetry of $\theta$, i.e. $V=V(\phi)$.%
\footnote{In general, top-down potentials can break the axionic shift symmetry and need not be single-valued on moduli space, see, e.g., \cite{Lust:2024aeg,Balavoine:2026jdj}.}
This shift symmetry is a symmetry of the cusp region, but not of the complete modular surface ${\cal M}_\tau={\mathbb H}^+/SL(2,{\mathbb Z})$, on which an exponential potential $V(\phi)$ is also not invariant under $SL(2,{\mathbb Z})$. The following analysis therefore applies to the cusp region rather than to the complete modular surface. We then make the following ansatz for the wave function, which takes into account the periodicity of $\theta$:
\begin{equation}
\Psi(a,\phi,\theta)= \sum_{n\in{\mathbb Z}}c_ne^{in\theta}\Psi_n(a,\phi)\, .
\label{ansatz}
\end{equation}
It follows that each $\Psi_n$ satisfies
\begin{equation}
\biggl({1\over 12 a}{d^2\over da^2}-{1\over 12a^2}{d\over da}-V_c^2\bigl(3ka-a^3V(\phi)\bigr)+{1\over a^3}\Bigl(-{1\over2}\partial^2_\phi+{\gamma\over2}\partial_\phi+{n^2\over 2f_0^2}e^{2\gamma|\phi |}\Bigr)\biggr)\Psi_n(a,\phi)=0\, .
\label{WDW3}
\end{equation}
Due to the factor $1/2$ in Eq.~\eqref{alapl}, the axionic term reproduces the classical axion kinetic energy, cf. Eq.~\eqref{eq:rho-theta}. In the remainder of this section we consider spatially flat slices with unit comoving volume, $k=0$ and $V_c=1$, so that $a^3$ is the spatial volume. For other values of $V_c$, the axionic momentum per unit comoving volume in Eq.~\eqref{eq:Jconserved} is $J=n/V_c$, and the semiclassical centrifugal energy density is obtained by replacing $n\rightarrow n/V_c$.

The exact WDW wave functions $\Psi_n(a,\phi)$ are difficult to determine. Instead, as in~\cite{Anchordoqui:2026nit}, we proceed with the minimization of the total potential, where we now have to include the correct dependence on the scale factor $a$. The total potential contains a geometric part from the Laplacian and the contribution of the actual potential $V(\phi)$, where we have multiplied both terms by an overall factor $a^3$:
\begin{equation}
V_{\rm total}(\phi)={V_{\rm geom}(\phi)}+a^6V(\phi)\, .
\end{equation}
For simplicity, we consider an exponential runaway potential $V(\phi)=V_0e^{-\alpha\phi}$ as in Eq.~\eqref{eq:general-rapidturn}. Adding the geometric potential, the total potential becomes
\begin{equation}
V_{\rm total}={n^2\over 2f_0^2}e^{2\gamma\phi}+a^6V_0e^{-\alpha\phi}\, .
\label{Vtotal}
\end{equation}
Note that if $\phi$ is identified with the dilaton and $\theta$ with the RR axion in the axio-dilaton, the geometric potential, which originates from the axion kinetic term, scales as $n^2g_s^{-2}$ and corresponds to a tree-level potential.

The total potential is minimized at
\begin{equation}
\phi_{{\rm min},n}={6\log a\over \alpha(1+\beta)}+{1\over \alpha(1+\beta)}\log\biggl({2f_0^2V_0\over \beta n^2}\biggr)\, ,\qquad\beta={2\gamma\over\alpha}\, .
\label{phimin}
\end{equation}
The value of the total potential at this minimum is
\begin{equation}
V_{{\rm min},n}(a)=a^{6\beta\over 1+\beta}\biggl({n^2\over 2f_0^2}\biggr)^{1\over 1+\beta}\biggl({V_{0}\over \beta}\biggr)^{\beta\over 1+\beta}\bigl(1+\beta\bigr)\, .
\label{Vmin}
\end{equation}
One can scale out a factor $a^6$, and then one obtains an effective energy density of the form
\begin{equation}
\Lambda_{{\rm eff},n}(a)=a^{-6}V_{{\rm min},n}(a)=a^{-{6\over 1+\beta}}\Lambda_{0,n}\, ,
\label{cceffective}
\end{equation}
with
\begin{equation}
\Lambda_{0,n}=\biggl({n^2\over 2f_0^2}\biggr)^{1\over 1+\beta}\biggl({V_0\over \beta}\biggr)^{\beta\over 1+\beta}\bigl(1+\beta\bigr)\, .
\label{Lambdaeff}
\end{equation}
In the factor $(1+\beta)$ appearing in $V_{{\rm min},n}(a)$ and $\Lambda_{0,n}$, the first term comes from the kinetic energy, whereas the second term, $\beta$, is the contribution from the potential energy (see also \cite{Krippendorf:2018tei} for a similar effect). After the minimization, both terms in the potential scale with the same power of $a$. For small $a$, i.e. in the early universe, the effective energy density is large compared to its value at $a=1$. Furthermore, the effective energy density grows in the limit $n\rightarrow\infty$. On the other hand, $\Lambda_{0,n}$ becomes small in the limit $n/f_0\rightarrow0$.

Let us now express $\Lambda_{0,n}$ in terms of $\phi_{{\rm min},n}$ to see how the effective energy density depends on the value of the scalar field. For simplicity, we set the scale factor to $a=1$ in Eq.~\eqref{phimin}. Then we obtain
\begin{equation}
\Lambda_{0,n}={V_0\over \beta}(1+\beta)e^{-\alpha\phi_{{\rm min},n}}\, .
\end{equation}
For relatively large $\beta$, which is the phenomenologically interesting case (see below), this relation further simplifies to
\begin{equation}
\Lambda_{0,n}\simeq V_0e^{-\alpha\phi_{{\rm min},n}}\, .
\end{equation}
These two relations show that the effective energy density becomes small for large field excursions in agreement with the AdS distance conjecture~\cite{Lust:2019zwm}.
However, as explained above, the limit of small $\Lambda_{0,n}$ and large $\phi_{{\rm min},n}$ requires a small value of $n/f_0$. This means that the quantum effects, which lead to the tracking of a moving minimum of $\phi$ and to a finite effective energy density, are nevertheless strongly suppressed for small $\Lambda_{0,n}$, and one is almost in the classical regime. 

Next, it is instructive to compute the equation of state parameter $w$, defined as
\begin{equation}
w={P\over \rho}={E_{\rm kin}-V\over E_{\rm kin}+V}\, ,
\end{equation}
where $P$ is the pressure, $\rho$ is the energy density of the system, and $E_{\rm kin}=a^{-6}V_{\rm geom}$ is the axionic kinetic energy density. Using $V=\beta E_{\rm kin}$ at the minimum, we obtain
\begin{equation}
w={1-\beta\over 1+\beta}\, .
\label{wus}
\end{equation}
For $\beta>1$ the equation of state parameter $w$ is negative, and for $\beta>2$ one has $w<-{1\over 3}$, i.e. accelerated expansion.\footnote{For recent studies of an axio-dilaton coupling~\cite{Cicoli:2020cfj,Burgess:2021obw,Cicoli:2023opf} to potentially  explain DESI-BAO data,  see~\cite{Shiu:2026edl,Anchordoqui:2026hys,Ramos-Hamud:2026wdx,Ramos-Hamud:2026aub,Yang:2026qon}.} Since $a$ increases with the expansion of the universe, the effective energy density \eqref{cceffective} decreases in time. At the same time, the value $\phi_{{\rm min},n}$ of the modulus field also increases in time, i.e. with increasing scale factor $a$, see Eq.~\eqref{phimin}. We emphasize again that the state found here corresponds to an excited state of the moduli space Hamiltonian with $n\neq0$. For $n=0$, the minimized potential vanishes, $\Lambda_{0,0}=0$, and the scalar field is driven to its runaway value, $\phi_{\rm min}=\infty$. This value is only the infimum of the minimized potential and not the energy of a normalizable quantum state, since the radial kinetic energy is positive, see Sec.~\ref{sec:localized}.

The scaling $\Lambda_{{\rm eff},n}\propto a^{-6/(1+\beta)}$ implies $a\propto t^{(1+\beta)/3}$, which together with Eq.~\eqref{wus} coincides with the exact axion-supported scaling solution of Sec.~\ref{sec:scaling-solutions} in $d=4$. In $d$ dimensions, $a^6\rightarrow a^{2(d-1)}$, and one finds $a\propto t^{(1+\beta)/(d-1)}$ and the same $w$. The minimization neglects the saxion kinetic energy, which on the exact solution is the fraction $\Omega_{\dot\phi}=6/(\alpha^2(1+\beta)^2)$ of the energy density in four dimensions, so its contribution to the energy budget is negligible when $\alpha^2(1+\beta)^2\gg6$. In particular, the minimization gives $\rho_\theta/V=1/\beta$ instead of the exact value ${\cal R}_{\rm AS}$. Since the exact solution satisfies $V=\beta(\rho_\theta+\dot\phi^2/2)$, the former is a good approximation only under the stronger condition $\alpha^2(1+\beta)\gg6$. On the other hand, $w=(1-\beta)/(1+\beta)$ is exact, since $w=1-2\Omega_V$ depends only on $\Omega_V=\beta/(1+\beta)$. The minimization therefore describes the late-time axion-supported regime discussed in Sec.~\ref{sec:transition}. For trajectories that pass through the tracker regime, $\rho_\theta/V$ initially grows approximately according to Eq.~\eqref{eq:R-growth}, while $w$ remains approximately at its tracker value as long as the axion contribution is negligible. During the subsequent transition, $w$ approaches its axion-supported value. The steep-potential route reaches the same attractor without an intermediate tracker phase.

Finally, note that we integrate out the field $\phi$ by setting it to its instantaneous minimum $\phi_{{\rm min},n}$, which yields the effective dependence on the scale factor $a$. This is a consistent approximation only if the mass of $\phi$ at its minimum is large compared to the Hubble parameter, $m_\phi\gg H$, with $H = \sqrt{\Lambda_{{\rm eff},n}/3}$ the Hubble parameter of the total effective energy density \eqref{cceffective}, and if the minimum moves sufficiently slowly. In this case $\phi$ is heavy and adiabatically tracks its moving minimum. However, if $m_\phi\lesssim H$, then $\phi$ could be displaced from its minimum, e.g. by quantum fluctuations or by its initial velocity, so it would not track its minimum. Since $\phi$ is canonically normalized, its mass is given by $m_\phi^2=a^{-6}\partial_\phi^2 V_{\rm total}$. Using the minimization condition for $\phi_{{\rm min},n}$, we find
\begin{equation}
{m_\phi^2\over H^2} = {3\, \partial_\phi^2 V_{\rm total}\over V_{\rm total}}=3\alpha^2\beta\, .
\end{equation}
Hence, the hierarchy $m_\phi\gg H$ requires $3\alpha^2\beta\gg1$, whereas the saxion kinetic energy only negligibly affects the balance of forces if $\alpha^2(1+\beta)\gg6$, as discussed above. Both conditions are satisfied for sufficiently large $\beta$.

\subsection{Quantum effects of the cusp}
\label{sec:localized}

As discussed in Sec.~\ref{sec:scaling}, axionic directions are expected to develop cusp-like geometries near infinite-distance limits \cite{Ooguri:2006in,Lanza:2021eft,Aoufia:2026ztl}. Locally, one therefore expects cusp-like metrics of the form
\begin{equation}
ds^2=d\phi^2+f_0^2e^{-2\gamma\phi}d\theta^2\, ,
\label{eq:cusp-general}
\end{equation}
where $\phi\rightarrow+\infty$ corresponds to the infinite-distance limit. This is the $\phi>0$ branch of the metric \eqref{2dimmetric}.

An important consequence of the cusp geometry is that a wave function localized in this region necessarily has a non-zero expectation value of the moduli-space Laplacian. For the metric \eqref{eq:cusp-general}, one has
\begin{equation}
-\Delta_{\cal M}=-\partial_\phi^2+\gamma\partial_\phi-\frac{1}{f_0^2}e^{2\gamma\phi}\partial_\theta^2\, .
\end{equation}
Expanding in axion momentum eigenstates,
\begin{equation}
\psi(\phi,\theta)=\sum_{n\in\mathbb Z}e^{in\theta}\psi_n(\phi)\, ,
\end{equation}
the operator in the $n$-th sector becomes
\begin{equation}
-\Delta_n=-\partial_\phi^2+\gamma\partial_\phi+\frac{n^2}{f_0^2}e^{2\gamma\phi}\, .
\end{equation}
The first-derivative term originates from the shrinking volume measure of the cusp. Defining $\psi_n(\phi)=e^{\gamma\phi/2}\chi_n(\phi)$, such that $\chi_n$ has the standard flat radial norm, gives
\begin{equation}
-\Delta_n\longrightarrow-\partial_\phi^2+\frac{\gamma^2}{4}+\frac{n^2}{f_0^2}e^{2\gamma\phi}\, .
\label{eq:cusp-schrodinger}
\end{equation}
Thus, even in the shift-invariant sector $n=0$, the asymptotic continuum begins at the positive threshold
\begin{equation}
\lambda_{\rm cusp}=\frac{\gamma^2}{4}\, .
\label{eq:cusp-gap}
\end{equation}
For $\gamma=1$ this reproduces the threshold $1/4$ of the modular spectrum \eqref{Ealpha}. We therefore conclude that, whenever there is a cusp-like asymptotic limit, the continuous spectrum associated with it starts at a positive threshold. More precisely, since $-\partial_\phi^2$ is non-negative and the remaining terms in Eq.~\eqref{eq:cusp-schrodinger} are bounded from below by $\lambda_{\rm cusp}$, any wave packet supported inside the cusp satisfies $\langle-\Delta_{\cal M}\rangle\geq\lambda_{\rm cusp}$. This is a statement about the asymptotic region and does not exclude eigenvalues below $\lambda_{\rm cusp}$ in the complete moduli space, such as the constant zero mode of a finite-volume moduli space. 

The quantization of the axionic momentum relies on the finiteness of the spatial volume. As in Sec.~\ref{sec:nonvanishingV}, we consider spatially flat slices with unit comoving volume, $V_c=1$, so that $a^{d-1}$ is the spatial volume in $d$-dimensional Planck units, which we assume to be large enough for the effective action to be a good approximation, for example $a>l_s$ in string theory. The momentum canonically conjugate to the homogeneous mode of $\theta$ is then the conserved axionic momentum \eqref{eq:Jconserved},
\begin{equation}
\pi_\theta=a^{d-1}f(\phi)^2\dot\theta=J\, ,
\label{eq:pi-theta}
\end{equation}
and integrating the commutator \eqref{commutatora3} over the spatial slice gives
\begin{equation}
\bigl[\theta,f(\phi)^2\dot\theta\bigr]=\frac{i}{a^{d-1}}\, .
\label{eq:zero-mode-commutator}
\end{equation}
Since $\theta\simeq\theta+2\pi$, the eigenvalues of $\pi_\theta=-i\partial_\theta$ are quantized, $J=n\in\mathbb Z$. For non-compact spatial slices, $J$ would instead be a momentum density and could take any value.

A semiclassical cosmology requires all moduli, including the axion, to take approximately definite values. The wave function therefore cannot be an eigenstate of $\pi_\theta$, which is completely delocalized in $\theta$, and in particular it cannot lie in the shift-invariant sector $n=0$. Instead, it must be a superposition of axionic momentum eigenstates, as in Eq.~\eqref{ansatz}, with support on $n\neq0$. Quantitatively, let the axion wave function be supported within an interval of length $\Delta\theta<2\pi$. The standard deviation of $\theta$ is then at most $\Delta\theta/2$, and the uncertainty relation implies $\Delta n\geq1/\Delta\theta$, so that
\begin{equation}
\langle n^2\rangle\geq\frac{1}{\Delta\theta^2}\, ,\qquad \bigl\langle\bigl(f(\phi)^2\dot\theta\bigr)^2\bigr\rangle=\frac{\langle n^2\rangle}{a^{2(d-1)}}\geq\frac{1}{\Delta\theta^2a^{2(d-1)}}\, .
\label{eq:n2-bound}
\end{equation}
Even if the classical axion momentum vanishes, $\langle n\rangle=0$, a localized axion therefore carries the kinetic energy density
\begin{equation}
\langle\rho_\theta\rangle=\frac{\langle n^2\rangle}{2f(\phi)^2a^{2(d-1)}}\geq\frac{1}{2\Delta\theta^2f(\phi)^2a^{2(d-1)}}\, .
\label{eq:rho-theta-quantum}
\end{equation}
These quantities vanish in the large-volume limit $a\rightarrow\infty$, in which the commutator \eqref{eq:zero-mode-commutator} vanishes and quantum effects die off, but they are non-zero for any finite spatial volume.

This has an immediate cosmological consequence. In four spacetime dimensions, the homogeneous matter Hamiltonian \eqref{alapl} contains
\begin{equation}
{\cal H}_{\rm matt}\supset-\frac{1}{2a^3}\Delta_{\cal M}\, ,
\end{equation}
which gives the comoving energy $\lambda_\psi/(2a^3)$, with $\lambda_\psi\equiv\langle\psi|-\Delta_{\cal M}|\psi\rangle$, and hence the energy density
\begin{equation}
\rho_{\rm q}(a)=\frac{\lambda_\psi}{2a^6}\, ,
\label{eq:universal-a6}
\end{equation}
in agreement with Eq.~\eqref{lambdaeff6} for $V_c=1$ and $E_{\rm matt}\equiv\lambda_\psi/2$. For a wave packet localized deep inside the cusp, one has $\lambda_\psi\geq\lambda_{\rm cusp}$. In the absence of a potential, $\lambda_\psi$ is constant. In general, however, the state evolves and $\lambda_\psi$ depends on $a$, as we will see below for its axionic part. In $d$ spacetime dimensions, the corresponding scaling is
\begin{equation}
\rho_{\rm q}(a)\sim\frac{\lambda_\psi}{2a^{2(d-1)}}\, .
\end{equation}
The axionic part of this energy is given by Eq.~\eqref{eq:rho-theta-quantum}. The explicit factor $a^{-2(d-1)}$ is the same volume suppression found above, whereas the exponentially shrinking axion decay constant can compensate this dilution as the saxion evolves. This is precisely the mechanism behind the axion-supported scaling solutions discussed in Sec.~\ref{sec:scaling-solutions}, see also Refs.~\cite{Christodoulidis:2019jsx,Revello:2023hro,Brown:2017hyper,Bjorkmo:2019rapid}.

Since, in the cusp region, the potential does not depend on $\theta$, the axionic momentum is conserved, and the WDW equation \eqref{WDW3} does not mix different momentum sectors. Moreover, the saxion and the scale factor couple to the axion only through $n^2$, via the axion kinetic energy in Eq.~\eqref{WDW3}. In the semiclassical regime, in which the saxion and the scale factor follow classical trajectories sourced by the expectation value of the axion kinetic energy, the cosmology is therefore described by the classical solutions of Sec.~\ref{sec:scaling}, with the conserved axionic momentum replaced by its root-mean-square value,
\begin{equation}
J\rightarrow n_{\rm rms}\equiv\sqrt{\langle n^2\rangle}\geq\frac{1}{\Delta\theta}\, ,
\label{eq:nrms}
\end{equation}
where we used Eq.~\eqref{eq:n2-bound}. While the expectation value $\langle n\rangle$ of the axionic momentum can vanish, $n_{\rm rms}$ cannot, since a localized axion requires a superposition of different momentum sectors. Different momentum sectors follow different classical trajectories and may eventually separate into distinct branches. If the $n=0$ sector separates, the resulting branch is an eigenstate of $\pi_\theta$ that is completely delocalized in $\theta$, and therefore does not describe a semiclassical solution with a classical axion. Assuming that the axion has a classical value, we must therefore focus on a branch with non-zero $n_{\rm rms}$, which is localized in $\theta$ and hence satisfies Eq.~\eqref{eq:nrms}. Once the $n=0$ sector has separated, every remaining component has $|n|\geq1$, so that in addition $n_{\rm rms}\geq1$. Since $\langle n^2\rangle$ is conserved within a branch, the initial fraction of the energy density in the axion kinetic energy, Eq.~\eqref{eq:initial-fractions}, is bounded from below,
\begin{equation}
\epsilon_\theta=\frac{n_{\rm rms}^2}{2f(\phi_i)^2a_i^{2(d-1)}\rho_i}\geq\frac{1}{2\Delta\theta^2f(\phi_i)^2a_i^{2(d-1)}\rho_i}\, ,
\label{eq:epsilon-theta-min}
\end{equation}
where $f(\phi_i)$, $a_i^{d-1}$, and $\rho_i$ are the axion decay constant, the spatial volume, and the energy density at the initial time. Classically, one could take the conserved angular momentum arbitrarily small and postpone this transition indefinitely. Quantum mechanically, $\epsilon_\theta$ cannot vanish for a localized axion. As shown in Sec.~\ref{sec:transition}, whenever the axion-supported scaling solution exists, i.e. $(d-2)\alpha^2(1+\beta)>4(d-1)$, any non-zero axion momentum eventually becomes important, and the system is driven to the axion-supported scaling attractor along each of the routes and for each of the initial conditions discussed there. Assuming that all moduli, including the axion, take semiclassical values, and that the effective description remains valid, the future evolution is therefore unavoidably driven by the axion-supported scaling cosmology after a finite time. Inserting Eq.~\eqref{eq:epsilon-theta-min} into the upper bound \eqref{eq:NRT-bound}, the number of $e$-folds before the axionic momentum becomes important satisfies
\begin{equation}
N_{\rm AS}\lesssim\begin{cases}\displaystyle\frac{1}{\lambda_R}\log\Bigl(2{\cal R}_{\rm AS}\Delta\theta^2f(\phi^i)^2a_i^{2(d-1)}\rho_i\Bigr)+\frac{\lambda_R+\lambda_K}{\lambda_R\,\mu_V}\log\Bigl(\frac{1}{\epsilon_V}\Bigr)\, , & (d-2)\alpha^2<4(d-1)\, ,\\[3mm] \displaystyle\frac{1}{2(d-1)}\Bigl[\log\Bigl(\frac{1}{\epsilon_V}\Bigr)+\frac{1}{\beta}\log\Bigl(2\Delta\theta^2f(\phi^i)^2a_i^{2(d-1)}\rho_i\Bigr)\Bigr]\, , & (d-2)\alpha^2>4(d-1)\, ,\end{cases}
\label{eq:NRT-quantum}
\end{equation}
where ${\cal R}_{\rm AS}$, $\lambda_R$, $\lambda_K$, $\mu_V$, and $\beta$ are given in terms of $d$, $\alpha$, and $\gamma$ in Eq.~\eqref{eq:NRT-bound-coefficients}. As in Eq.~\eqref{eq:NRT-bound}, this bound retains the leading logarithmic dependence on the initial energy fractions. Since $\Delta\theta<2\pi$, the right-hand side is in turn bounded by its value at $\Delta\theta=2\pi$, which gives the maximal delay for any localized axion. The bound grows only logarithmically with the initial spatial volume $a_i^{d-1}$, and a sharper localization of the axion leads to an earlier onset. This estimate depends on the initial position of the wave packet, as expected, but the important qualitative difference from the classical theory is that the root-mean-square angular momentum cannot be taken continuously to zero.

\section{A new inequality for axions}
\label{sec:TCC}

We next discuss the implications of the axion-supported scaling solutions for the trans-Planckian censorship conjecture (TCC) \cite{Bedroya:2019snp}. As shown in Sec.~\ref{sec:localized}, whenever the axion-supported scaling solution exists, it eventually drives the evolution of any semiclassical state in which the axion is localized. Its compatibility with the TCC therefore constrains the asymptotic behavior of the axion decay constant and of the scalar potential, and leads to a new inequality between their exponential rates $\gamma$ and $\alpha$. We then contrast this inequality with the strong de Sitter conjecture, test it in top-down limits, and relate it to light EFT strings.

\subsection{TCC for axion-supported scaling cosmologies}
\label{sec:TCC-RT}

Recall from Sec.~\ref{sec:scaling-solutions} that, for an axion decay constant $f=f_0e^{-\gamma\phi}$ and a scalar potential $V=V_0e^{-\alpha\phi}$, the axion-supported scaling solution has $a(t)\propto t^p$ with $p=(1+\beta)/(d-1)$ and $\beta=2\gamma/\alpha$, see Eq.~\eqref{eq:p-beta-d}, and that it exists whenever the condition \eqref{eq:rapidturn-existence-d} holds. Acceleration requires $p>1$, i.e. $\beta>d-2$, see Eq.~\eqref{eq:acceleration-d} and Ref.~\cite{Revello:2023hro}. If such an accelerating scaling solution remained under perturbative control for an arbitrarily long time, it would violate the TCC. Indeed,
\begin{equation}
aH\propto t^{p-1}\, ,
\end{equation}
and hence for $p>1$ the comoving Hubble radius $(aH)^{-1}$ shrinks without bound, so that modes which were initially shorter than the quantum-gravity cutoff length eventually cross the Hubble radius. Therefore, an asymptotic axion-supported scaling solution which remains inside the regime of validity of the EFT must satisfy
\begin{equation}
\boxed{2\gamma\leq(d-2)\alpha}\, ,
\label{eq:TCC-rapidturn}
\end{equation}
or equivalently $\beta\leq d-2$. In four dimensions this reduces to $\gamma\leq\alpha$. An accelerating solution with the opposite inequality is not necessarily inconsistent if it is only transient, but it must end before the TCC bound is violated. Since the axion-supported scaling solution is the late-time attractor whenever it exists (see Sec.~\ref{sec:transition}), this can only happen if the exponential behavior of the axion decay constant or of the scalar potential, or the validity of the effective description itself, breaks down. Hence, the inequality \eqref{eq:TCC-rapidturn} must hold in any asymptotic limit in which the axion-supported scaling solution exists and the effective description remains valid.

The axion-supported scaling solutions also show that there is a distinction between conditions stated directly in terms of the scalar potential, such as the strong de Sitter conjecture \cite{Rudelius:2021oaz,Rudelius:2021azq}
\begin{equation}
\frac{|\nabla V|}{V}\geq\frac{2}{\sqrt{d-2}}\, ,
\label{eq:strong-dSC}
\end{equation}
and conditions stated in terms of the solutions, such as the TCC, even if we restrict our attention to future infinity. For a single field with an exponential potential, the two agree asymptotically, since the tracker solution accelerates if and only if $\alpha<2/\sqrt{d-2}$. In the cusp, however, the axion-supported scaling solution accelerates whenever $2\gamma>(d-2)\alpha$, independently of the value of $\alpha$. In particular, in the range of parameters
\begin{equation}
\frac{2}{\sqrt{d-2}}\leq\alpha<\frac{2\gamma}{d-2}\, ,
\label{eq:strong-dSC-TCC}
\end{equation}
the scalar potential satisfies the strong de Sitter conjecture, whereas the axion-supported scaling solution exists, since $(d-2)\alpha(\alpha+2\gamma)>(d-1)(d-2)\alpha^2\geq4(d-1)$, is the late-time attractor, and violates the TCC. Therefore, the TCC has a prediction about cusps, Eq.~\eqref{eq:TCC-rapidturn}, which does not follow from conditions on the scalar potential alone.

\subsection{Top-down tests}
\label{sec:top-down}

Interestingly, the inequality \eqref{eq:TCC-rapidturn} is obeyed in the standard controlled infinite-distance limits of string theory. This is a highly non-trivial test of the TCC, since the exponential rates $\alpha$ and $\gamma$ are not arbitrary in quantum gravity. One can further constrain $\alpha$ and $\gamma$, and in this way derive constraints on
\begin{equation}
w={\alpha-2\gamma\over\alpha+2\gamma}\, .
\label{walpha}
\end{equation}
In particular, the taxonomy rules, which are closely related to the Emergent String Conjecture (ESC), force the exponential rates of the potential and of the axion decay constant, viewed as vectors on moduli space, to be lattice-valued. When the non-compact modulus is a radion or a dilaton, as in the examples below, this forces $\alpha$ and $\gamma$, and hence $w$, to be discrete. Specifically, it was argued in \cite{Aoufia:2026ztl} that $\gamma$ is required to lie on the lattice of $\alpha$-vectors for branes with a zero-dimensional worldvolume ($p=0$), i.e. instantons, where this lattice was defined in \cite{Etheredge:2025ahf}. In addition, it was demonstrated in \cite{Etheredge:2026cha} that $\alpha$ is also lattice-valued and lies on the lattice of $\alpha$-vectors for spacetime-filling branes with a $d$-dimensional worldvolume ($p=d$), also defined in \cite{Etheredge:2025ahf}. These two observations imply constraints on $w$. Using the terminology of \cite{Etheredge:2025ahf}, let
\begin{equation}
\alpha=\alpha_d(P_d)
\end{equation}
be the $\alpha$-vector of a codimension-0 brane, and let
\begin{equation}
\gamma=\alpha_0(P_0)
\end{equation}
be the $\alpha$-vector of an instanton. Since the $\alpha$-vectors of Ref.~\cite{Etheredge:2025ahf} are defined as $-\nabla\log T$, with $T$ the tension, and the action of an instanton that couples to the axion scales as $1/f$, the $\alpha$-vector of this instanton is $-\gamma$, so that $P_0$ is the negative of its label. Then we have
\begin{equation}
\alpha\pm 2\gamma=\alpha_d(P_d\pm2 P_0)\, .
\end{equation}
That is, $\alpha \pm2 \gamma$ lies on the lattice of codimension-0 branes, at the site $P=P_d\pm 2P_0$. In particular, if the non-compact modulus is a radion corresponding to decompactification from $d$ to $D$ dimensions, we find that $w$ is discretized and equal to
\begin{equation}
w={d(D-2)-(P_d-2P_0)(d-2)\over d(D-2)-(P_d+2P_0)(d-2)}\, .
\end{equation}
Instead, if the non-compact modulus is the dilaton of a $d$-dimensional theory, we find that $w$ is also discretized and equal to
\begin{equation}
w={2d+(d-2)(d-P_d+2P_0)\over2d+(d-2)(d-P_d-2P_0)}\, .
\end{equation}
In summary, the recent taxonomy rules for potentials and axion decay constants imply that the values of $w$ are discretized. They depend on the integers $d$ and $D$, which refer to spacetime dimensions, and on the integers $P_d$ and $P_0$, which control the decay rates of the potential and of the axion decay constant.

The TCC inequality \eqref{eq:TCC-rapidturn}, which is equivalent to $w\geq-(d-3)/(d-1)$, is therefore a non-trivial condition on the lattice sites that can be realized in asymptotic limits. We now verify it in the standard decompactification and perturbative emergent-string limits.

Consider first a decompactification from $d$ to $D=d+k$ dimensions, in which the saxion $\phi$ is the canonically normalized radion. An axion obtained by reducing a $q$-form gauge field on a $q$-cycle has asymptotic decay constant
\begin{equation}
f^2\sim\exp\biggl(-2q\sqrt{\frac{d-2}{k(D-2)}}\,\phi\biggr)\, ,
\end{equation}
and hence
\begin{equation}
\gamma=q\sqrt{\frac{d-2}{k(D-2)}}\, .
\label{eq:gamma-decomp}
\end{equation}
On the other hand, a $d$-dimensional potential descending from a local $D$-dimensional operator of Weyl weight $\delta\geq0$ has exponential slope \cite{Etheredge:2026cha}
\begin{equation}
\alpha=\sqrt{\frac{d-2}{k(D-2)}}\left(\frac{2k}{d-2}+\delta\right)\, .
\label{eq:alpha-decomp}
\end{equation}
Using $q\leq k$, one immediately finds
\begin{equation}
(d-2)\alpha-2\gamma=\sqrt{\frac{d-2}{k(D-2)}}\,\Bigl[2(k-q)+(d-2)\delta\Bigr]\geq0\, .
\label{eq:TCC-decomp-check}
\end{equation}
Thus the TCC inequality is automatically satisfied throughout this class of decompactification limits. Saturation requires both $q=k$ and $\delta=0$. In this case, however, $\alpha^2(d-2)=4k/(D-2)<4$, so that the existence condition \eqref{eq:rapidturn-existence-d} with $\beta=d-2$ is violated and no axion-supported branch exists.

A similar result holds in standard perturbative emergent-string limits. Let the saxion $\phi\rightarrow+\infty$ now denote the canonically normalized weak-coupling direction. The leading closed-string axions have decay constants of the form \cite{Aoufia:2026ztl}
\begin{equation}
f\sim e^{-\gamma\phi}\, ,\qquad \gamma=\frac{\ell}{2}\sqrt{d-2}\, ,
\end{equation}
with $\ell=1$ for ordinary RR axions and $\ell=2$ for dual NS axions in the standard closed-string sectors. A potential term generated directly in $d$ dimensions and scaling as $g_s^{k_g}$ in string frame, with $k_g\geq-2$ ($k_g=-2$ at tree level), has asymptotic slope \cite{Etheredge:2026cha}
\begin{equation}
\alpha=\frac{d+\frac{k_g}{2}(d-2)}{\sqrt{d-2}}\geq\frac{2}{\sqrt{d-2}}\, .
\end{equation}
Therefore,
\begin{equation}
2\gamma=\ell\sqrt{d-2}\leq2\sqrt{d-2}\leq(d-2)\alpha\, .
\label{eq:TCC-string-check}
\end{equation}
The inequality is strict for the ordinary RR axions. The limiting case $\ell=2$ together with a tree-level term $k_g=-2$ saturates the TCC bound. In this case the existence condition \eqref{eq:rapidturn-existence-d} is saturated as well, so the axion-supported branch merges with the tracker solution, which gives the marginal coasting solution $a(t)\propto t$, rather than accelerated expansion.

As a simple four-dimensional example, the standard perturbative string values are
\begin{equation}
\gamma_{\rm RR}=\frac{1}{\sqrt2}\, ,\qquad \gamma_{\rm NS}=\sqrt2\, ,
\end{equation}
whereas the shallowest exponential allowed by the above scaling, which arises at tree level, has
\begin{equation}
\alpha_{\rm min}=\sqrt2\, .
\end{equation}
Thus the RR case satisfies the inequality strictly, while the dual NS case saturates it.

These observations suggest a useful connection between the asymptotic geometry of moduli space, the unavoidable quantum kinetic contribution associated with localized wave functions, and the TCC. The cusp geometry forces localized semiclassical states to carry non-zero moduli-space kinetic energy. Axionic components of this energy can support scaling cosmologies, but the known top-down asymptotic limits obey precisely the inequality required to prevent an eternally accelerating axion-supported branch. It would be interesting to understand whether the inequality \eqref{eq:TCC-rapidturn} follows more generally from the taxonomy of axion decay constants and scalar potentials in quantum gravity \cite{Etheredge:2025ahf,Aoufia:2026ztl,Etheredge:2026cha}.

\subsection{Light EFT strings and asymptotic accelerated expansion}
\label{sec:EFT-strings}

The inequality \eqref{eq:TCC-rapidturn} has a remarkable counterpart in recent studies of asymptotic accelerated expansion in four-dimensional ${\cal N}=1$ theories. In Ref.~\cite{Calderon-Infante:2022nxb}, the possibility of asymptotic accelerated expansion from the runaway flux potentials of Calabi-Yau compactifications, classified in Ref.~\cite{Grimm:2019ixq}, was analyzed. Two candidate trajectories with asymptotic accelerated expansion, which arise from competing terms in the flux potential, could not be ruled out there: a Type IIB limit at large complex structure and weak coupling with $\alpha=\sqrt{2/7}$, and an F-theory limit away from weak coupling with $\alpha=2/\sqrt5$. Both involve only the complex-structure moduli and assume that the K\"ahler moduli are stabilized. In Ref.~\cite{Collazuol:2026xuz}, building on the EFT strings of Refs.~\cite{Lanza:2020qmt,Lanza:2021eft} and their relation to towers of light states \cite{Grieco:2025bjy,Grieco:2026yip,Etheredge:2026jun}, both trajectories were ruled out, because an EFT string becomes light compared to the Hubble scale. Along the first trajectory the tension scale of the EFT string decays twice as fast as the Hubble scale, whereas along the second it decays at exactly the same rate, and in both cases the species scale also drops below the Hubble scale. It was also shown there that the quantum obstructions to ${\cal N}=1$ infinite-distance limits of Refs.~\cite{Kaufmann:2026fli,Kaufmann:2026mha,Kaufmann:2026tsy}, which change the asymptotic form of the K\"ahler potential, can resolve this issue and also prevent the asymptotic acceleration, unless the quantum-corrected volumes can be stabilized while the complex-structure moduli continue to run. The condition used in Ref.~\cite{Collazuol:2026xuz}, $H\lesssim{\cal T}^{1/2}$ with ${\cal T}$ the tension of the EFT string, makes no reference to axions. However, along a trajectory the ratio of the exponential decay rates of the EFT string scale and of the Hubble scale, which Ref.~\cite{Collazuol:2026xuz} requires to be smaller than one, is precisely $\gamma/\alpha$ for the axion dual to the EFT string, so that this condition becomes our inequality \eqref{eq:TCC-rapidturn} in four dimensions, $\gamma\leq\alpha$. In particular, the two trajectories above have $\gamma=2\alpha$ and $\gamma=\alpha$, respectively, and the marginal case $\gamma=\alpha$ was also regarded as problematic in Ref.~\cite{Collazuol:2026xuz}.

This is not a coincidence. In $d$ spacetime dimensions, the EFT string generalizes to the $(d-3)$-brane magnetically charged under the axion $\theta\simeq\theta+2\pi$, whose tension is
\begin{equation}
{\cal T}_{d-3}=\frac{2\pi}{\gamma}fM_p^{d-2}\, .
\label{eq:Tbrane}
\end{equation}
For $d=4$ this reproduces the tension of the BPS EFT strings of four-dimensional ${\cal N}=1$ theories \cite{Lanza:2021eft}. For example, for a saxion $s$ with K\"ahler potential $K=-n\log s$, whose axion partner has unit period, one finds $\gamma=\sqrt{2/n}$ and ${\cal T}_1=nM_p^2/(2s)$. Since the brane cannot be resolved within the EFT, its tension scale must lie above the species scale $\Lambda_s$, which in turn must lie above the Hubble scale, $H\lesssim\Lambda_s\lesssim{\cal T}_{d-3}^{1/(d-2)}$ \cite{Collazuol:2026xuz}. A bottom-up argument that the tension of a brane must lie above the species scale, based on gravitational amplitudes and black hole thermodynamics, was given in Ref.~\cite{Bedroya:2024ubj}, see also Ref.~\cite{Grieco:2025bjy} for EFT strings. In particular, the EFT breaks down unless ${\cal T}_{d-3}\gtrsim H^{d-2}$. Along the axion-supported scaling solution, and more generally along any trajectory on which $H^2$ scales like the potential, one has ${\cal T}_{d-3}\propto e^{-\gamma\phi}$ and $H^{d-2}\propto e^{-(d-2)\alpha\phi/2}$, so that this condition reproduces exactly the inequality \eqref{eq:TCC-rapidturn}, $2\gamma\leq(d-2)\alpha$.

We therefore find two completely independent origins for the inequality \eqref{eq:TCC-rapidturn}. First, the brane magnetically charged under the axion must have a tension scale above the Hubble scale for the EFT not to break down. Second, the TCC must hold for the axion-supported scaling solutions in the cusp. Asymptotic limits that violate the bound are therefore likely obstructed. Related EFT constraints on how the scalar potential and the masses of towers of light states can scale at infinite distance were obtained from their cosmological dynamics in Ref.~\cite{Casas:2024oak}.

It is remarkable that, in all the ${\cal N}=1$ flux compactifications analyzed in Ref.~\cite{Collazuol:2026xuz}, every flow along which the scalar potential alone gives accelerated expansion has an axion with $\gamma\geq\alpha$, whereas every flow that satisfies the strong de Sitter conjecture \eqref{eq:strong-dSC} strictly has $\gamma<\alpha$ for all axions. More precisely, in four-dimensional ${\cal N}=1$ supergravity one cannot use the axion-supported scaling solutions to obtain eternal accelerated expansion from a potential that would otherwise not give eternal accelerated expansion. This is consistent with $\gamma=\sqrt{2/n}\leq\sqrt2$ for $n\geq1$, so that the acceleration condition $\gamma>\alpha$ of the axion-supported scaling solution requires $\alpha<\sqrt2$, for which the potential alone already violates the strong de Sitter conjecture and gives accelerated expansion. It is also in line with the observation that rapid-turn inflation in supergravity is rare \cite{Aragam:2021scu}.

\section{Summary}
\label{sec:summary}

We have revisited the axion-supported scaling cosmologies of Refs.~\cite{Christodoulidis:2019jsx,Revello:2023hro} in the cusp geometries that arise in top-down theories, which give rise to faster expansion for the same scalar potential due to the non-trivial coupling between the axion and the saxion. Axionic infinite-distance limits are expected to develop cusp-like geometries in which the axion decay constant decreases exponentially with the saxion, and we considered a classical potential $V(\phi)$ given by a simple exponential of $\phi$, as is typical for a large class of scalar potentials in string theory (see \cite{Etheredge:2026cha} for a classification of these potentials). The non-zero axionic momentum gives an energy density proportional to $a^{-6}$ ($a^{-2(d-1)}$ in $d$ spacetime dimensions), while the exponentially decreasing axion decay constant enhances this contribution as the saxion moves towards the cusp. These two effects can balance the exponential potential and lead to axion-supported scaling solutions, which for a large turn rate are the well-known rapid-turn solutions. For an asymptotic axion-saxion metric with axion decay constant $f\sim e^{-\gamma\phi}$ and a scalar potential $V\sim e^{-\alpha\phi}$ in $d$ spacetime dimensions, the axion-supported scaling solution has $a(t)\propto t^{(1+\beta)/(d-1)}$, with $\beta=2\gamma/\alpha$, and equation of state
\begin{equation}
w=\frac{\alpha-2\gamma}{\alpha+2\gamma}\, .
\end{equation}
Whenever it exists, it expands faster than the solutions driven by the same potential in the absence of the axion.

Our main new result is that, quantum mechanically, these solutions are universal when spatial slices are compact. Employing the well-known WDW equation of quantum gravity and using previous results from moduli space quantum mechanics, we have demonstrated how the geometry of moduli space can make quantum effects unavoidable for semiclassical cosmological states. Coupling the quantized scalar fields of the internal geometry to the spacetime metric, we find that the kinetic energy associated with the moduli-space Laplacian is suppressed by the factor $a^{-3}$ ($a^{-(d-1)}$ in $d$ spacetime dimensions), where $a$ is the spatial scale factor, see Eq.~\eqref{alapl}. In particular, in four spacetime dimensions a non-zero expectation value of the moduli-space Laplacian gives rise to an effective energy density that scales as $a^{-6}$ for a fixed expectation value, see Eq.~\eqref{eq:universal-a6}. Therefore, even if these quantum effects are small at large spatial volume, they become increasingly important towards the past, i.e. at smaller values of the scale factor.

We have investigated the WDW equation, the associated wave functions, and their energies in a toy model based on the two-dimensional moduli space ${\cal M}_\tau={\mathbb H}^+/SL(2,{\mathbb Z})$ with two scalar fields, the saxion $\phi$ and the axion $\theta$. This moduli space is quite generic, as it describes the modular invariant kinetic energy of these two scalar fields and typically arises at the infinite-distance boundary of moduli space. In the presence of a potential, we restricted the analysis to its cusp region, in which the axionic shift symmetry is approximately restored. The Laplacian in such a cusp has the positive asymptotic threshold \eqref{eq:cusp-gap}. Although the complete moduli space may contain a constant zero mode, this state is completely delocalized and does not describe a semiclassical configuration in which the moduli take approximately definite values. A wave packet localized in the cusp therefore necessarily contains Laplacian modes with non-zero eigenvalue. If the axion itself is localized, the wave packet must in addition contain non-zero axionic momentum modes. The new quantum-mechanical observation is that, on compact spatial slices, the axionic momentum is quantized, and that a semiclassical wave function in which the axion is localized necessarily has a non-zero root-mean-square axionic momentum, see Eq.~\eqref{eq:nrms}. Hence, its contribution cannot be taken continuously to zero and, whenever $(d-2)\alpha(\alpha+2\gamma)>4(d-1)$, i.e. whenever the axion-supported scaling solution exists, the future evolution is unavoidably driven by it after a finite number of $e$-folds, see Eq.~\eqref{eq:NRT-quantum}.

We showed that, starting from a phase dominated by the saxion kinetic energy, the axion kinetic energy, or the potential, the system reaches the axion-supported scaling solution after a number of $e$-folds that depends only logarithmically on the initial fractions of the energy density in the potential and in the axion kinetic energy, see Eq.~\eqref{eq:NRT-bound}. Together with the quantization of the axion momentum, this gives an upper bound on the time-scale before the axion-supported scaling solution takes over, see Eq.~\eqref{eq:NRT-quantum}.

Finally, we asked whether the cusp region can give rise to accelerated expansion, which requires $2\gamma>(d-2)\alpha$. We do not find such examples in top-down constructions, and indeed the TCC requires a non-trivial bound on the cusps. Since the axion-supported scaling solution is the late-time attractor whenever it exists, an asymptotically accelerating axion-supported solution would violate the TCC unless the effective description breaks down. The TCC therefore requires the new inequality for axions
\begin{equation}
2\gamma\leq(d-2)\alpha\, .
\end{equation}
This is a prediction about cusps that does not follow from conditions on the scalar potential alone, since the strong de Sitter conjecture can be satisfied in part of the parameter space that violates this inequality, see Eq.~\eqref{eq:strong-dSC-TCC}. Remarkably, this inequality is obeyed in the standard controlled decompactification and perturbative emergent-string limits of string theory, see Eqs.~\eqref{eq:TCC-decomp-check} and \eqref{eq:TCC-string-check}. It also follows independently from requiring that the brane magnetically charged under the axion remains heavier than the Hubble scale, which reproduces the criterion recently used to rule out candidate accelerating trajectories in four-dimensional ${\cal N}=1$ flux compactifications \cite{Collazuol:2026xuz}, see Eq.~\eqref{eq:Tbrane}. This suggests a direct connection between the asymptotic geometry of moduli space, the quantum-mechanical localization of semiclassical states, axion-supported scaling cosmologies, and the constraints imposed by quantum gravity. It would be interesting to see whether more involved string compactifications, with metrics involving several axion-saxion fields and multi-field potentials and refined taxonomy rules, can allow for accelerated expansion with $w<-1/3$.

{\makeatletter\let\addcontentsline\@gobblethree\makeatother\section*{Acknowledgements}}

We are grateful for conversations with Christian Aoufia, Bernardo Fraiman, Jacob McNamara, Sanjay Raman, Matthew Reece, Tom Rudelius, and Alexander Stewart. The work of L.A.A. is supported by the U.S. National Science Foundation (NSF Grant PHY-2412679). He also extends his appreciation to the Harvard Swampland Initiative for their warm hospitality and for providing a stimulating environment for productive discussions. The work of A.B. is supported in part by the U.S. Department of Energy under grant number DE-FG02-91ER40671 and by the Simons Foundation under grant number 654561, and A.B. acknowledges the support of Grant 63749 from the John Templeton Foundation. The opinions expressed in this publication are those of the author(s) and do not necessarily reflect the views of the John Templeton Foundation or the National Science Foundation. The work of D.L. is supported by the German-Israel-Project (DIP) on ``Holography and the Swampland.''
A.W. is partially supported by the Deutsche Forschungsgemeinschaft under Germany’s Excellence Strategy - EXC 2121 ``Quantum Universe'' - 390833306, by the Deutsche Forschungsgemeinschaft through a German-Israeli Project Cooperation (DIP) grant ``Holography and the Swampland'', and by the Deutsche Forschungsgemeinschaft through the Collaborative Research Center SFB1624  ``Higher Structures, Moduli Spaces, and Integrability.''
\vskip1cm

\appendix

\section{Spacetime dynamics of non-canonical field commutators}
\label{sec:appA}

In this Appendix we compare the field-momentum and field-velocity commutation relations. Consider a minimally coupled scalar field $\phi$ in a spatially flat FLRW background. Omitting the spatial-gradient terms, which do not affect the canonical momentum, the Lagrangian density $\mathscr{L}$, defined relative to the coordinate volume element $dt~d^3x$, is
\begin{equation}
\mathscr{L}=a^{3}\left(\frac{1}{2}\dot{\phi }^{2} - V(\phi )\right)  \, .
\end{equation}
The canonically conjugate momentum density $\pi(\mathbf{x})$ is obtained by differentiating this Lagrangian density with respect to the field velocity $\dot \phi$,
\begin{equation}
\pi (\mathbf{x})=\frac{\partial \mathscr{L}}{\partial \dot{\phi}(\mathbf{x})}=a^{3}\dot{\phi }(\mathbf{x}) \, .
\end{equation}
When moving from classical field theory to quantum field theory, we promote fields to operators. The equal-time canonical quantization condition demands that a field and its own conjugate momentum obey the fundamental Poisson-bracket-to-commutator mapping
\begin{equation}
[\phi (\mathbf{x}),\pi (\mathbf{y})]=i\delta^{(3)}(\mathbf{x}-\mathbf{y}) \, .
\label{phipi}
\end{equation}
The equal-time commutator $[\phi(\mathbf{x}), \pi(\mathbf{y})]$ remains completely unchanged in an FLRW background and satisfies the standard canonical relation \eqref{phipi}. This is because canonical commutation relations are a fundamental postulate of quantum mechanics and quantum field theory. They are imposed at the level of the phase space variables, independently of the choice of spacetime metric or scale factor $a(x^0)$.

While the $\phi$-$\pi$ commutator is invariant, the commutator between the field and its velocity $\dot \phi$ absorbs the dependence on the scale factor,
\begin{equation}
[\phi (\mathbf{x}),a^{3}\dot{\phi }(\mathbf{y})]=i\delta ^{(3)}(\mathbf{x}-\mathbf{y})\, ,
\end{equation}
and so
\begin{equation}
[\phi (\mathbf{x}),\dot{\phi }(\mathbf{y})]=\frac{i}{a^{3}}\delta ^{(3)}(\mathbf{x}-\mathbf{y}) \, .
\label{phidotphi}
\end{equation}
Since $a^{-3}\delta^{(3)}(\mathbf{x}-\mathbf{y})$ is the delta function with respect to the physical volume, the commutator between the field $\phi$ and its velocity $\dot\phi$, averaged over a fixed comoving region, dilutes dynamically as $a^{-3}$ as the universe expands, in agreement with Eq.~\eqref{commutatora3}.

The difference between $[\phi,\pi]$ and $[\phi,\dot \phi]$ lies in canonical invariance versus spacetime dynamics. The momentum commutator is a fixed law of quantum mechanics, while the velocity commutator changes as the universe expands. On the one hand, the momentum commutator $[\phi, \pi]$ pairs a coordinate with its canonical partner, $\pi = \partial \mathscr{L}/\partial \dot{\phi}$. It is a foundational rule of quantum field theory, analogous to $[x, p] = i\hbar$ in standard quantum mechanics. Therefore, it is independent of the spacetime geometry, i.e. of gravity, the expansion of space, and the metric. It must always equal $i \delta^{(3)}(\mathbf{x} - \mathbf{y})$ to preserve the laws of quantum mechanics. On the other hand, the velocity commutator $[\phi, \dot{\phi}]$ pairs a coordinate with its rate of change in cosmic time. In an FLRW background, the expanding volume factor $a^{3}$ changes how velocity maps to momentum, $\pi = a^3 \dot{\phi}$. As the universe expands, a fixed comoving volume contains more physical space. Hence, for fields averaged over a fixed comoving volume, the commutator between the value of a field and its velocity dilutes away.

In physics, two variables are canonically conjugate only if they form a fundamental coordinate pair $(q, p)$ in Hamiltonian mechanics. This requires their Poisson bracket to equal one, or equivalently their commutator to equal $i$ (or $i$ times a delta function for fields). Note that velocity and momentum are identical only if the mass, here the factor $a^3$, is exactly one. Because the FLRW universe has a changing scale factor $a(x^0)$, the ``mass'' of the field effectively changes over time.

If two variables are true canonical conjugates, their commutator cannot depend on the dynamics of spacetime. It must be an absolute constant. As derived in Eq.~\eqref{phidotphi}, the commutator between the field and its velocity changes dynamically. Because the right-hand side of Eq.~\eqref{phidotphi} dilutes as the universe expands, $\phi$ and $\dot{\phi }$ fail the definition of canonically conjugate variables. They are simply a field and its time derivative.

In summary, in an FLRW background, the canonical equal-time field-momentum commutator satisfies the standard identity \eqref{phipi}. This relation is a fundamental axiom of quantization and remains unaffected by cosmic expansion, whereas the field-velocity commutator scales dynamically as $a^{-3}$.

\section{Taxonomy and commutation rules in $0+1$-dimensional quantum gravity}
\label{sec:appB}

As explained in \cite{Anchordoqui:2025izb,Anchordoqui:2026nit}, the ESC-related taxonomy rules \cite{Castellano:2023stg,Castellano:2023jjt,Etheredge:2024tok, Etheredge:2025ahf, Aoufia:2026ztl, Etheredge:2026cha} can be equivalently formulated as commutation relations in moduli space quantum mechanics. For example, the CRV pattern \cite{Castellano:2023stg,Castellano:2023jjt} relating the species number $N_s(\phi)$ and the mass scale $m_\text{lightest}$ of the lightest tower in $d$-dimensional Minkowski spacetime takes the form
\begin{align}
	\nabla \log m_\text{lightest}\cdot \nabla \log N_s=-1\, .\label{CRV}
\end{align}
Using the quantum mechanical relation
\begin{equation}
\left[ F(\phi),\dot G(\phi)\right]=i\, \nabla F(\phi)\cdot \nabla G(\phi)\, ,\label{fhcom}
\end{equation}
the CRV pattern translates into the commutation relation
\begin{align}
	\left[\log N_s,\frac d{dt}\log m_\text{lightest} \right]=-i\, .\label{fhcom2}
\end{align}
Furthermore, other taxonomic dot-product conditions of \cite{Etheredge:2024tok, Etheredge:2025ahf, Aoufia:2026ztl, Etheredge:2026cha} also can be expressed as commutators, using equation \eqref{fhcom}.

In a curved four-dimensional FLRW spacetime, the effective field-space metric of the homogeneous mechanics, $(a(t))^3g_{ij}$, also contains the volume factor $(a(t))^3$, see Eq.~\eqref{action2}. While the CRV pattern \eqref{CRV} with respect to the moduli space metric $g_{ij}$ is unchanged, the contraction with the inverse of the effective field-space metric leads to an inverse volume factor,
\begin{align}
	(a(t))^{-3}g^{ij}\partial_i\log m_\text{lightest}\,\partial_j\log N_s=-(a(t))^{-3}\, .\label{CRV1}
\end{align}
Consequently, the corresponding commutation relation, which involves the effective field-space metric through the field-velocity commutator, contains the same volume factor. Taking the operators $\log N_s$ and $\frac{d}{dt}\log m_\text{lightest}$ at two different spatial points $\vec x$ and $\vec y$, we expect
\begin{align}
	\left[\log N_s(\vec x),\frac d{dt}\log m_\text{lightest} (\vec y)\right]=-i   \delta^{(3)}(\vec x -\vec y)  (a(t))^{-3 }\, .\label{fhcom3}
\end{align}

This can also be checked by a direct computation of the commutator. For that, we consider the case of a single scalar field $\varphi$ with~\cite{Anchordoqui:2025izb}
\begin{equation}
N_s=\varphi\, ,\quad m_\text{lightest}=1/\varphi^\alpha\, ,\quad   g_{\varphi\varphi}=\alpha/\varphi^2  \, ,
\end{equation}
which satisfies Eq.~\eqref{CRV}, and we evaluate the left-hand side of Eq.~\eqref{fhcom3} using Eq.~\eqref{commutatora3}:
\begin{eqnarray}
\lbrack \log N_s(\vec x), {d\over dt}\log m_\text{lightest}(\vec y)\rbrack&=&{1\over \varphi}\lbrack \varphi(\vec x),{d\over dt}\log m_\text{lightest}(\vec y)\rbrack \nonumber\\
&=&{1\over \varphi}{\partial_\varphi m_\text{lightest}\over m_\text{lightest}}\lbrack \varphi(\vec x),\dot\varphi(\vec y)\rbrack\nonumber\\
&=& -(a(t))^{-3 } g_{\varphi\varphi}^{-1}{\alpha\over \varphi^2}\lbrack \varphi(\vec x), \pi(\vec y) \rbrack\nonumber\\
&=&-(a(t))^{-3 } \lbrack \varphi(\vec x),\pi(\vec y) \rbrack=-i \delta^{(3)}(\vec x -\vec y)(a(t))^{-3 } \, .\label{fhcom7}
\end{eqnarray}
Note that $d(\log m_{\rm lightest})/dt$ scales with $\dot\varphi$. Thus, even though neither the mass of the Kaluza-Klein modes nor $m_{\rm lightest}$ redshifts away, the commutator between $\log N_s$ and the cosmic-time derivative of $\log m_{\rm lightest}$, for operators averaged over a fixed comoving volume, decreases as $a^{-3}$. This behavior resembles that of the commutation relation between a field and its velocity, see Appendix~\ref{sec:appA} for details.

In closing, we note that all other commutators discussed in~\cite{Anchordoqui:2026nit} acquire the same additional volume dependence under time evolution.


\vskip0.5cm

\begin{thebibliography}{99}



\bibitem{Vafa:2005ui}
C.~Vafa,
{\color{rossoCP3}The String landscape and the swampland},
[arXiv:hep-th/0509212 [hep-th]].

\bibitem{Ooguri:2006in}
H.~Ooguri and C.~Vafa,
{\color{rossoCP3}On the Geometry of the String Landscape and the Swampland},
Nucl. Phys. B \textbf{766}, 21-33 (2007)
[arXiv:hep-th/0605264 [hep-th]].

\bibitem{Lee:2019wij}
S.J. Lee, W. Lerche, and T. Weigand,
{\color{rossoCP3}Emergent strings from infinite distance limits},
JHEP \textbf{02}, 190 (2022)
[arXiv:1910.01135 [hep-th]].

\bibitem{Etheredge:2022opl}
M. Etheredge, B. Heidenreich, S. Kaya, Y. Qiu, and T. Rudelius,
{\color{rossoCP3}Sharpening the Distance Conjecture in diverse dimensions},
JHEP \textbf{12}, 114 (2022)
[arXiv:2206.04063 [hep-th]].

\bibitem{Castellano:2023stg}
A.~Castellano, I.~Ruiz and I.~Valenzuela,
{\color{rossoCP3}  Universal pattern in quantum gravity at infinite distance},
Phys. Rev. Lett. \textbf{132}, no.18, 181601 (2024)
[arXiv:2311.01501 [hep-th]].

\bibitem{Castellano:2023jjt}
A.~Castellano, I.~Ruiz and I.~Valenzuela,
{\color{rossoCP3} Stringy evidence for a universal pattern at infinite distance},
JHEP \textbf{06}, 037 (2024)
[arXiv:2311.01536 [hep-th]].

\bibitem{Etheredge:2024tok}
M.~Etheredge, B.~Heidenreich, T.~Rudelius, I.~Ruiz and I.~Valenzuela,
{\color{rossoCP3}  Taxonomy of infinite distance limits},
JHEP \textbf{03}, 213 (2025)
[arXiv:2405.20332 [hep-th]].

\bibitem{Etheredge:2025ahf}
M.~Etheredge,
{\color{rossoCP3}Taxonomy of branes in infinite distance limits},
JHEP \textbf{10} (2025), 200
[arXiv:2505.10615 [hep-th]].

\bibitem{Aoufia:2026ztl}
C.~Aoufia, M.~Etheredge, B.~Fraiman, S.~Raman and A.~Stewart,
{\color{rossoCP3}Moduli-Space Laplacians, Asymptotic Geometry, and the Emergent String Conjecture},
[arXiv:2607.20603 [hep-th]].

\bibitem{Etheredge:2026cha}
M.~Etheredge, D.~L{\"u}st and T.~Rudelius,
{\color{rossoCP3}Taxonomy of Potentials in Asymptotic Limits},
[arXiv:2609.03001 [hep-th]].

\bibitem{Lanza:2021eft}
S.~Lanza, F.~Marchesano, L.~Martucci and I.~Valenzuela,
{\color{rossoCP3}The EFT stringy viewpoint on large distances},
JHEP \textbf{09} (2021), 197
[arXiv:2104.05726 [hep-th]].

\bibitem{Christodoulidis:2019jsx}
P.~Christodoulidis, D.~Roest and E.~I.~Sfakianakis,
{\color{rossoCP3}Scaling attractors in multi-field inflation},
JCAP \textbf{12} (2019), 059
[arXiv:1903.06116 [hep-th]].

\bibitem{Revello:2023hro}
F.~Revello,
{\color{rossoCP3}Attractive (s)axions: cosmological trackers at the boundary of moduli space},
JHEP \textbf{05} (2024), 037
[arXiv:2311.12429 [hep-th]].

\bibitem{Brown:2017hyper}
A.~R.~Brown,
{\color{rossoCP3}Hyperbolic Inflation},
Phys. Rev. Lett. \textbf{121} (2018) no.25, 251601
[arXiv:1705.03023 [hep-th]].

\bibitem{Bjorkmo:2019rapid}
T.~Bjorkmo,
{\color{rossoCP3}Rapid-Turn Inflationary Attractors},
Phys. Rev. Lett. \textbf{122} (2019) no.25, 251301
[arXiv:1902.10529 [hep-th]].

\bibitem{Mizuno:2017idt}
S.~Mizuno and S.~Mukohyama,
{\color{rossoCP3}Primordial perturbations from inflation with a hyperbolic field-space},
Phys. Rev. D \textbf{96} (2017) no.10, 103533
[arXiv:1707.05125 [hep-th]].

\bibitem{Bjorkmo:2019aev}
T.~Bjorkmo and M.~C.~D.~Marsh,
{\color{rossoCP3}Hyperinflation generalised: from its attractor mechanism to its tension with the `swampland conditions'},
JHEP \textbf{04} (2019), 172
[arXiv:1901.08603 [hep-th]].

\bibitem{Seo:2024qzf}
M.~S.~Seo,
{\color{rossoCP3}Asymptotic Behavior of Saxion-Axion System in Stringy Quintessence Model},
Fortsch. Phys. \textbf{72} (2024) no.11, 2400112
doi:10.1002/prop.202400112
[arXiv:2403.07307 [hep-th]].

\bibitem{Grimm:2025cpq}
T.~W.~Grimm, D.~van de Heisteeg and F.~Revello,
{\color{rossoCP3}Axion-scalar systems and dynamical distances},
JHEP \textbf{06} (2026), 184
[arXiv:2510.12879 [hep-th]].

\bibitem{Anchordoqui:2025izb}
L.~A.~Anchordoqui, D.~L\"ust and S.~L{\"u}st,
{\color{rossoCP3}Species Quantum Mechanics},
Phys. Rev. D \textbf{113}, no.2, 026021 (2026)
[arXiv:2510.25846 [hep-th]].

\bibitem{Anchordoqui:2026nit}
L.~Anchordoqui, M.~Etheredge and D.~L\"ust,
{\color{rossoCP3}Moduli Space Quantum Mechanics},
[arXiv:2603.06795 [hep-th]].

\bibitem{Wheeler:1968iap}
  J.~A.~Wheeler,
{\color{rossoCP3}Superspace and the nature of quantum geometrodynamics}, in Battelle
Rencontres, 1967 Lectures in Mathematics and Physics
(Eds. C. M. DeWitt and J. A. Wheeler, W. A. Benjamin, New York, 1968)
pp.~242--307; Adv. Ser. Astrophys. Cosmol. \textbf{3}, 27-92 (1987)

\bibitem{DeWitt:1967yk}
B.~S.~DeWitt,
{\color{rossoCP3}Quantum Theory of Gravity 1: The canonical theory},
Phys. Rev. \textbf{160}, 1113-1148 (1967)

\bibitem{Gielen:2023hsg}
S.~Gielen,
{\color{rossoCP3}Quantum Cosmology},
[arXiv:2309.01790 [gr-qc]].

\bibitem{Achucarro:2018vey}
A.~Ach\'ucarro and G.~A.~Palma,
{\color{rossoCP3}The string swampland constraints require multi-field inflation},
JCAP \textbf{02} (2019), 041
[arXiv:1807.04390 [hep-th]].

\bibitem{Akrami:2020zfz}
Y.~Akrami, M.~Sasaki, A.~R.~Solomon and V.~Vardanyan,
{\color{rossoCP3}Multi-field dark energy: Cosmic acceleration on a steep potential},
Phys. Lett. B \textbf{819} (2021), 136427
[arXiv:2008.13660 [astro-ph.CO]].

\bibitem{Rudelius:2021oaz}
T.~Rudelius,
{\color{rossoCP3}Dimensional reduction and (Anti) de Sitter bounds},
JHEP \textbf{08} (2021), 041
[arXiv:2101.11617 [hep-th]].

\bibitem{Rudelius:2021azq}
T.~Rudelius,
{\color{rossoCP3}Asymptotic observables and the swampland},
Phys. Rev. D \textbf{104} (2021) no.12, 126023
[arXiv:2106.09026 [hep-th]].

\bibitem{Bedroya:2019snp}
A.~Bedroya and C.~Vafa,
{\color{rossoCP3} Trans-Planckian Censorship and the Swampland},
JHEP \textbf{09}, 123 (2020)
[arXiv:1909.11063 [hep-th]].

\bibitem{Bedroya:2022tbh}
A.~Bedroya,
{\color{rossoCP3}Holographic origin of TCC and the distance conjecture},
JHEP \textbf{06} (2024), 016
doi:10.1007/JHEP06(2024)016
[arXiv:2211.09128 [hep-th]].

\bibitem{Bedroya:2024zta}
A.~Bedroya, Q.~Lu and P.~J.~Steinhardt,
{\color{rossoCP3}TCC in the interior of moduli space and its implications for the string landscape and cosmology},
JHEP \textbf{08} (2025), 007
doi:10.1007/JHEP08(2025)007
[arXiv:2407.08793 [hep-th]].

\bibitem{vandeHeisteeg:2023uxj}
D.~van de Heisteeg, C.~Vafa, M.~Wiesner and D.~H.~Wu,
{\color{rossoCP3}Bounds on field range for slowly varying positive potentials},
JHEP \textbf{02} (2024), 175
doi:10.1007/JHEP02(2024)175
[arXiv:2305.07701 [hep-th]].

\bibitem{Goheer:2002vf}
N.~Goheer, M.~Kleban and L.~Susskind,
{\color{rossoCP3}The Trouble with de Sitter space},
JHEP \textbf{07}, 056 (2003)
doi:10.1088/1126-6708/2003/07/056
[arXiv:hep-th/0212209 [hep-th]].

\bibitem{Dvali:2014gua}
G.~Dvali and C.~Gomez,
{\color{rossoCP3}Quantum Exclusion of Positive Cosmological Constant?},
Annalen Phys. \textbf{528} (2016), 68-73
[arXiv:1412.8077 [hep-th]].

\bibitem{Maltz:2016iaw}
J.~Maltz and L.~Susskind,
{\color{rossoCP3}de Sitter Space as a Resonance},
Phys. Rev. Lett. \textbf{118}, no.10, 101602 (2017)
doi:10.1103/PhysRevLett.118.101602
[arXiv:1611.00360 [hep-th]].

\bibitem{Dvali:2017eba}
G.~Dvali, C.~Gomez and S.~Zell,
{\color{rossoCP3}Quantum Break-Time of de Sitter},
JCAP \textbf{06} (2017), 028
[arXiv:1701.08776 [hep-th]].

\bibitem{Dvali:2018fqu}
G.~Dvali and C.~Gomez,
{\color{rossoCP3}On Exclusion of Positive Cosmological Constant},
Fortsch. Phys. \textbf{67} (2019) no.1-2, 1800092
[arXiv:1806.10877 [hep-th]].

\bibitem{Dvali:2018jhn}
G.~Dvali, C.~Gomez and S.~Zell,
{\color{rossoCP3}Quantum Breaking Bound on de Sitter and Swampland},
Fortsch. Phys. \textbf{67} (2019) no.1-2, 1800094
[arXiv:1810.11002 [hep-th]].

\bibitem{Calderon-Infante:2022nxb}
J.~Calder\'on-Infante, I.~Ruiz and I.~Valenzuela,
{\color{rossoCP3}Asymptotic accelerated expansion in string theory and the Swampland},
JHEP \textbf{06} (2023), 129
[arXiv:2209.11821 [hep-th]].

\bibitem{Collazuol:2026xuz}
V.~Collazuol, A.~Grieco and I.~Valenzuela,
{\color{rossoCP3}Light EFT Strings and Asymptotic Accelerated Expansion},
[arXiv:2610.24062 [hep-th]].

\bibitem{Halliwell:1986ja}
J.~J.~Halliwell,
{\color{rossoCP3}Scalar Fields in Cosmology with an Exponential Potential},
Phys. Lett. B \textbf{185} (1987), 341

\bibitem{Copeland:1997et}
E.~J.~Copeland, A.~R.~Liddle and D.~Wands,
{\color{rossoCP3}Exponential potentials and cosmological scaling solutions},
Phys. Rev. D \textbf{57} (1998), 4686-4690
[arXiv:gr-qc/9711068 [gr-qc]].

\bibitem{Misner:1967uu}
C.~W.~Misner,
{\color{rossoCP3}The Isotropy of the universe},
Astrophys. J. \textbf{151} (1968), 431-457

\bibitem{Gasperini:1996fu}
M.~Gasperini, J.~Maharana and G.~Veneziano,
{\color{rossoCP3}Graceful exit in quantum string cosmology},
Nucl. Phys. B \textbf{472} (1996), 349-360
[arXiv:hep-th/9602087 [hep-th]].

\bibitem{McNamara:2020uza}
J.~McNamara and C.~Vafa,
{\color{rossoCP3}Baby Universes, Holography, and the Swampland},
[arXiv:2004.06738 [hep-th]].

\bibitem{Harlow:2025pvj}
D.~Harlow, M.~Usatyuk and Y.~Zhao,
{\color{rossoCP3}Quantum mechanics and observers for gravity in a closed universe},
JHEP \textbf{02} (2026), 108
[arXiv:2501.02359 [hep-th]].

\bibitem{Vilenkin:1982de}
A.~Vilenkin,
{\color{rossoCP3}Creation of Universes from Nothing},
Phys. Lett. B \textbf{117} (1982), 25-28

\bibitem{Hartle:1983ai}
J.~B.~Hartle and S.~W.~Hawking,
{\color{rossoCP3}Wave Function of the Universe},
Phys. Rev. D \textbf{28} (1983), 2960-2975

\bibitem{Dvali:2007hz}
G.~Dvali,
{\color{rossoCP3}Black Holes and Large N Species Solution to the Hierarchy Problem},
Fortsch. Phys. \textbf{58} (2010), 528-536
[arXiv:0706.2050 [hep-th]].

\bibitem{Dvali:2007wp}
G.~Dvali and M.~Redi,
{\color{rossoCP3}Black Hole Bound on the Number of Species and Quantum Gravity at LHC},
Phys. Rev. D \textbf{77} (2008), 045027
[arXiv:0710.4344 [hep-th]].

\bibitem{Calderon-Infante:2023ler}
J.~Calder\'on-Infante, A.~Castellano, A.~Herr\'aez and L.~E.~Ib\'a\~nez,
{\color{rossoCP3}Entropy bounds and the species scale distance conjecture},
JHEP \textbf{01} (2024), 039
doi:10.1007/JHEP01(2024)039
[arXiv:2306.16450 [hep-th]].

\bibitem{vandeHeisteeg:2023ubh}
D.~van de Heisteeg, C.~Vafa and M.~Wiesner,
{\color{rossoCP3}Bounds on species scale and the distance conjecture},
Fortsch. Phys. \textbf{71} (2023), 2300143
doi:10.1002/prop.202300143
[arXiv:2303.13580 [hep-th]].

\bibitem{Long:2021jlv}
C.~Long, M.~Montero, C.~Vafa and I.~Valenzuela,
{\color{rossoCP3}The desert and the swampland},
JHEP \textbf{03} (2023), 109
doi:10.1007/JHEP03(2023)109
[arXiv:2112.11467 [hep-th]].

\bibitem{vandeHeisteeg:2022btw}
D.~van de Heisteeg, C.~Vafa, M.~Wiesner and D.~H.~Wu,
{\color{rossoCP3}Moduli-dependent species scale},
Beijing J. Pure Appl. Math. \textbf{1} (2024), 1
[arXiv:2212.06841 [hep-th]].



\bibitem{Cribiori:2022nke}
N.~Cribiori, D.~L{\"u}st and G.~Staudt,
{\color{rossoCP3}Black hole entropy and moduli-dependent species scale},
Phys. Lett. B \textbf{844} (2023), 138113
[arXiv:2212.10286 [hep-th]].



\bibitem{vandeHeisteeg:2023dlw}
D.~van de Heisteeg, C.~Vafa, M.~Wiesner and D.~H.~Wu,
{\color{rossoCP3}Species scale in diverse dimensions},
JHEP \textbf{05} (2024), 112
[arXiv:2310.07213 [hep-th]].

\bibitem{Green:1997tv}
M.~B.~Green and M.~Gutperle,
{\color{rossoCP3}Effects of D-instantons},
Nucl. Phys. B \textbf{498} (1997), 195-227
[arXiv:hep-th/9701093 [hep-th]].

\bibitem{Gibbons:1978ac}
G.~W.~Gibbons, S.~W.~Hawking and M.~J.~Perry,
{\color{rossoCP3}Path integrals and the indefiniteness of the gravitational action},
Nucl. Phys. B \textbf{138} (1978), 141-150.

\bibitem{Hohm:2019ccp}
O.~Hohm and B.~Zwiebach,
{\color{rossoCP3}Duality invariant cosmology to all orders in $\alpha'$},
Phys. Rev. D \textbf{100} (2019), 126011
[arXiv:1905.06963 [hep-th]].

\bibitem{Hohm:2019jgu}
O.~Hohm and B.~Zwiebach,
{\color{rossoCP3}Non-perturbative de Sitter vacua via $\alpha'$ corrections},
Int. J. Mod. Phys. D \textbf{28} (2019), 1943002
[arXiv:1905.06583 [hep-th]].

\bibitem{Codina:2021cxh}
T.~Codina, O.~Hohm and D.~Marques,
{\color{rossoCP3}General string cosmologies at order $\alpha'^3$},
Phys. Rev. D \textbf{104} (2021), 106007
[arXiv:2107.00053 [hep-th]].

\bibitem{DHoker:2022dxx}
E.~D'Hoker and J.~Kaidi,
{\color{rossoCP3}Lectures on modular forms and strings},
[arXiv:2208.07242 [hep-th]].


\bibitem{Lust:2024aeg}
S.~L{\"u}st,
{\color{rossoCP3}An index for flux vacua},
JHEP \textbf{01} (2026), 144
[arXiv:2405.04584 [hep-th]].

\bibitem{Balavoine:2026jdj}
P.~Balavoine and S.~L{\"u}st,
{\color{rossoCP3}Flux Vacua, Asymptotic Periods, and Relatively Anosov Representations},
[arXiv:2610.27077 [hep-th]].

\bibitem{Krippendorf:2018tei}
S.~Krippendorf, F.~Muia and F.~Quevedo,
{\color{rossoCP3}Moduli Stars},
JHEP \textbf{08} (2018), 070
[arXiv:1806.04690 [hep-th]].



\bibitem{Lust:2019zwm}
D.~L{\"u}st, E.~Palti and C.~Vafa,
{\color{rossoCP3}AdS and the Swampland},
Phys. Lett. B \textbf{797} (2019), 134867
[arXiv:1906.05225 [hep-th]].







\bibitem{Cicoli:2020cfj}
M.~Cicoli, G.~Dibitetto and F.~G.~Pedro,
{\color{rossoCP3}New accelerating solutions in late-time cosmology},
Phys. Rev. D \textbf{101}, no.10, 103524 (2020)
doi:10.1103/PhysRevD.101.103524
[arXiv:2002.02695 [gr-qc]].

\bibitem{Burgess:2021obw}
C.~P.~Burgess, D.~Dineen and F.~Quevedo,
{\color{rossoCP3}Yoga Dark Energy: natural relaxation and other dark implications of a supersymmetric gravity sector},
JCAP \textbf{03}, no.03, 064 (2022)
doi:10.1088/1475-7516/2022/03/064
[arXiv:2111.07286 [hep-th]].


\bibitem{Cicoli:2023opf}
M.~Cicoli, J.~P.~Conlon, A.~Maharana, S.~Parameswaran, F.~Quevedo and I.~Zavala,
 {\color{rossoCP3} String cosmology: From the early universe to today},
Phys. Rept. \textbf{1059}, 1-155 (2024)
doi:10.1016/j.physrep.2024.01.002
[arXiv:2303.04819 [hep-th]].


\bibitem{Shiu:2026edl}
G.~Shiu, F.~Tonioni and H.~V.~Tran,
{\color{rossoCP3} Bounding axion dark energy},
JCAP \textbf{09}, 066 (2026)
doi:10.1088/1475-7516/2026/09/066
[arXiv:2604.09141 [astro-ph.CO]].


\bibitem{Anchordoqui:2026hys}
L.~A.~Anchordoqui and D.~L\"ust,
{\color{rossoCP3} Breaking Free from the Swampland of Impossible Universes Through the DESI Portal},
Universe \textbf{12}, no.7, 216 (2026)
doi:10.3390/universe12070216
[arXiv:2605.10476 [astro-ph.CO]].

\bibitem{Ramos-Hamud:2026wdx}
M.~Ramos-Hamud, G.~Garc{\'\i}a-Arroyo, F.~Quevedo and J.~A.~V{\'a}zquez,
{\color{rossoCP3}Axio-Dilaton Dark Energy: A Dynamical Systems and Bayesian Inference Analysis},
[arXiv:2609.22539 [astro-ph.CO]].


\bibitem{Ramos-Hamud:2026aub}
M.~Ramos-Hamud,
{\color{rossoCP3}Spiralling into Dark Energy: Axio-Dilaton Attractors in Light of Current Data},
[arXiv:2610.34380 [hep-th]].


\bibitem{Yang:2026qon}
W.~Yang, S.~Zhang, S.~Pan, L.~A.~Anchordoqui and E.~Di Valentino, 
{\color{rossoCP3} Exploring the Limits: Neutrino Mass and Interacting Dark Sector in the DESI Era},
[arXiv:2610.33729 [hep-ph]].
  

\bibitem{Grimm:2019ixq}
T.~W.~Grimm, C.~Li and I.~Valenzuela,
{\color{rossoCP3}Asymptotic Flux Compactifications and the Swampland},
JHEP \textbf{06} (2020), 009
[erratum: JHEP \textbf{01} (2021), 007]
[arXiv:1910.09549 [hep-th]].

\bibitem{Lanza:2020qmt}
S.~Lanza, F.~Marchesano, L.~Martucci and I.~Valenzuela,
{\color{rossoCP3}Swampland Conjectures for Strings and Membranes},
JHEP \textbf{02} (2021), 006
[arXiv:2006.15154 [hep-th]].

\bibitem{Grieco:2025bjy}
A.~Grieco, I.~Ruiz and I.~Valenzuela,
{\color{rossoCP3}EFT strings and dualities in 4d ${\cal N}=1$},
JHEP \textbf{06} (2026), 129
[arXiv:2504.16984 [hep-th]].

\bibitem{Grieco:2026yip}
A.~Grieco, I.~Ruiz and I.~Valenzuela,
{\color{rossoCP3}EFT (String) Tower Building},
[arXiv:2607.21700 [hep-th]].



\bibitem{Etheredge:2026jun}
M.~Etheredge, A.~Herr{\'a}ez, D.~L{\"u}st and L.~Melotti,
{\color{rossoCP3}Integral Scaling for EFT Strings from the Bottom-Up},
[arXiv:2607.22519 [hep-th]].







\bibitem{Kaufmann:2026fli}
L.~Kaufmann, J.~Monnee, T.~Weigand and M.~Wiesner,
{\color{rossoCP3}Quantum obstructions for ${\cal N}=1$ infinite distance limits. I. $g_s$ obstructions},
Phys. Rev. D \textbf{113} (2026) no.12, 126027
[arXiv:2603.12315 [hep-th]].

\bibitem{Kaufmann:2026mha}
L.~Kaufmann, J.~Monnee, T.~Weigand and M.~Wiesner,
{\color{rossoCP3}Quantum obstructions for ${\cal N}=1$ infinite distance limits. II. K\"ahler obstructions},
Phys. Rev. D \textbf{113} (2026) no.12, 126028
[arXiv:2603.13470 [hep-th]].

\bibitem{Kaufmann:2026tsy}
L.~Kaufmann, T.~Weigand and M.~Wiesner,
{\color{rossoCP3}On Quantum Obstructions in Type IIA Orientifolds},
[arXiv:2604.25988 [hep-th]].

\bibitem{Bedroya:2024ubj}
A.~Bedroya, R.~K.~Mishra and M.~Wiesner,
{\color{rossoCP3}Density of states, black holes and the Emergent String Conjecture},
JHEP \textbf{01} (2025), 144
[arXiv:2405.00083 [hep-th]].

\bibitem{Casas:2024oak}
G.~F.~Casas and I.~Ruiz,
{\color{rossoCP3}Cosmology of light towers and swampland constraints},
JHEP \textbf{12} (2024), 193
[arXiv:2409.08317 [hep-th]].

\bibitem{Aragam:2021scu}
V.~Aragam, R.~Chiovoloni, S.~Paban, R.~Rosati and I.~Zavala,
{\color{rossoCP3}Rapid-turn inflation in supergravity is rare and tachyonic},
JCAP \textbf{03} (2022) no.03, 002
[arXiv:2110.05516 [hep-th]].


\end{thebibliography}
\end{document}